# THE NEUTRINO
## Its past, present and future

Ivan V. Aničin
*Faculty of Physics, University of Belgrade, Belgrade,
Serbia and Montenegro*

The review consists of two parts. In the first part the critical points in the past, present and future of neutrino physics (nuclear, particle and astroparticle) are briefly reviewed. In the second part the contributions of Yugoslav physics to the physics of the neutrino are commented upon. The review is meant as a first reading for the newcomers to the field of neutrino physics.

**Table of contents**

**Introduction**

The neutrinos appear to constitute by number of species not less than one quarter of the particles which make the world, and even half of the stable ones. By number of particles in the Universe they are perhaps second only to photons. Yet, seventy years after we first suspected their existence, amazingly little is known about their detailed properties.

According to the views of the actual Standard Model of elementary particles and their interactions (the SM for short) all matter is built out of only 12 structureless fermions: 6 quarks and 6 leptons, which fall into left-handed weak isospin doublets of the three generations:

$$\begin{pmatrix} u \\ d' \end{pmatrix}_L \begin{pmatrix} c \\ s' \end{pmatrix}_L \begin{pmatrix} t \\ b' \end{pmatrix}_L \quad \begin{matrix} 2/3 \\ -1/3 \end{matrix}$$
$$\begin{pmatrix} \nu_e \\ e^- \end{pmatrix}_L \begin{pmatrix} \nu_\mu \\ \mu^- \end{pmatrix}_L \begin{pmatrix} \nu_\tau \\ \tau^- \end{pmatrix}_L \quad \begin{matrix} 0 \\ -1 \end{matrix} \qquad (1)$$
$$1st \quad 2nd \quad 3rd \quad Q$$

The leptons do not experience strong interactions while the universality of the weak interaction requires the mixing of strong interaction (or mass) quark eigenstates $d$, $s$, $b$ into the primed weak eigenstates:

$$\begin{pmatrix} d' \\ s' \\ b' \end{pmatrix} = \begin{pmatrix} U_{ud} & U_{us} & U_{ub} \\ U_{cd} & U_{cs} & U_{cb} \\ U_{td} & U_{ts} & U_{tb} \end{pmatrix} \begin{pmatrix} d \\ s \\ b \end{pmatrix}. \qquad (2)$$

where the matrix elements of the unitary CKM (Cabibbo-Kobayashi-Maskawa) transformation matrix are determined experimentally. To each doublet in (1) there corresponds a weak interaction vertex in which the two fermions are coupled to a charged W boson, always with the same weak coupling constant. The experimental width of the neutral Z boson resonance corresponds to three neutrino flavors almost exactly, what asserts that the scheme is complete, and is one of the cornerstones of the SM.

The masses of the neutrinos of all three flavors are, by Occam's razor, all assumed in the SM to exactly vanish, implying that the lepton families do not mix, and that the three lepton numbers are separately exactly conserved. The neutrinos are further assumed to be Dirac particles, the antineutrino being genuinely different from the neutrino of the same flavor.

Due to the extremely feeble interaction ability of the neutrinos, which are the only electrically neutral fundamental fermions and do not bind to the systems of other stable particles, much hard work went into establishing this picture, but whether the leptons indeed behave as assumed by the SM is not at all certain. Different alternative properties of the neutrino, and/or of the weak interaction, are not excluded by experimental evidence which we accumulated up to now.

In what follows we give an outline of the evolution of our knowledge of the neutrino, discuss where we stand today, and comment on some possible future developments. Relations of Yugoslav physics to some episodes of the physics of the neutrino are reviewed. The first part is the personal view of the events in which the author did not actively participate, while those described in the second part are known to the author in much greater detail.

References to historical work which has become common knowledge, which are available in most textbooks on nuclear and particle physics, are not explicitly cited in the general review. Complementary and more detailed information about the history of neutrino physics is to be found in many excellent extensive reviews and recollections on the subject, and references therein (1-8). Most of the subtleties of neutrino physics are probably best reviewed and systematized in Ref.25, while all but the latest numerical results are summarized in Ref. 13.

**A. GENERAL REVIEW OF NEUTRINO PHYSICS**

**A.1. Short history of the neutrino**

A.1.1. First epoch; 1930-1956
   *Neutrino is conjectured, put into a sound theoretical frame, and finally detected*

In the momentous late 1920's, even after the rise of both non-relativistic and relativistic quantum mechanics, the atoms were thought to have been composed only of protons and electrons. The electrons were supposed to exist both in the nucleus (the nuclear electrons, which were thought to be ejected in beta decay) as well as around it (the atomic electrons). The need for a certain, yet unseen nuclear particle, which had to carry energy and spin in the processes of nuclear beta decay together with the beta electron, became especially pressing when after some twenty years of hard experimenting the calorimetric measurements of the total energy released in the decay gave same values as the integrated continuous spectrum obtained by the magnetic spectrometry of the emitted beta radiation, and when this fell short of the Q value of the decay as obtained from mass measurements. To conserve electric charge and account for its obvious low interacting ability the particle had to be neutral, had to have negligible mass, and spin 1/2. With only two particles known at the time it took one Pauli to dare to postulate the existence of the third one, what he did in 1930, first rather timidly and later on with somewhat more self-confidence, in spite of Bohr's opposition, who rather believed in the violation of fundamental conservation laws in beta decay (though Bohr was prophetic; some conservation laws indeed turned out to be violated in beta decay!). Pauli supposed the particle, which he tentatively named the "neutron", to have just the properties which the experiments told that it had to have. The initial confusion around the thus contemplated particle was amplified by the long time felt lack of another neutral nuclear particle, the "zero-th element" of Rutherford, or the "neutral proton" of Majorana, but when such a particle was discovered in 1932 by Chadwick, the ambiguity was quickly resolved - the neutron of Chadwick and the neutron of Pauli were seen to be two different particles! The neutron of Chadwick, the genuine elementary particle and not the strongly bound system of a proton and an electron, was found to fit the nucleus ideally, and the nuclear force, later to become the strong interaction, which had to keep the protons and neutrons together, had to be devised. The nuclear electrons were needed no more (everybody was relieved), and the neutron of Pauli also did not fit into the nucleus any more. It took one Fermi to finally solve the riddle. In analogy with quantum electrodynamics, on which he wrote an encyclopedic review article two years earlier, Fermi, in 1933-34, postulated the existence of the fourth force of nature, later to become known as the weak interaction, which causes the nucleus of A nucleons and atomic number Z to change into its isobar of atomic number Z+1 while creating the beta electron and the Pauli neutron, which, to reduce confusion, he renamed into "neutrino", the "small neutron", in Italian. The theory reproduced the shape of the (allowed) beta spectra perfectly (though after some doubts, since the at that time best known beta spectrum of RaE did not behave well) and this was considered as a sufficiently strong, though indirect evidence, that the neutrino is as real as is the observable beta electron. It was clear that the neutrino experiences only this peculiar kind of interaction and ever since the story of the neutrino overlaps with that of the weak interaction. Fermi postulated the rest mass of the neutrino to be exactly null, and the interaction to be the point interaction of four participating fermions, or the two charged 4-currents (the nucleon and the lepton one), of vector transformation properties (this in analogy with electrodynamics, what some 25 years later turned out to have been a happy choice). The coupling constant thus defined became known as the Fermi constant, $G_F$. This pattern of the discovery of the neutrino was inspiring and was, with variations, repeated many times in the following development of particle physics.

In the same 1934 Bethe and Peierls estimated that the cross section for the neutrino to induce nuclear processes, e.g. to induce the reaction named the "inverse beta decay" (*mutatis mutandis*, in contemporary notation, from the beta decay of the neutron):

$$\bar{\nu} + p \rightarrow e^+ + n, \qquad (3)$$

should be so small, of the order of $10^{-43}$ cm$^2$ for an 1 MeV (anti)neutrino (though, at not too high energies it would increase with energy squared) that its mean free path in solid matter would be of the order of tens of light-years. This left little hope that neutrino will ever be explicitly seen and suggested that indirect evidence of its existence should be patiently accumulated. Miraculously, however, as it will turn out, after another twenty years of intensive experimental development, the neutrino will be directly observed.

One year later, in 1935, Yukawa reduced all the elementary interactions to the exchange of virtual bosons, including the idea that even the Fermi point interaction might actually be realized by the exchange of an extremely heavy intermediate vector boson, what would manifest only at energies higher than the boson mass. In this he would be proved right almost fifty years later.

In 1936 Gamow and Teller made important addition to the original Fermi theory, who considered only the case of the lepton pair carrying off the zero angular momentum. They considered the case of beta decay where the nucleus may change its spin and the lepton pair is emitted with parallel spins. The interaction which leads to such a "GT" transition is different from that which leads to the "Fermi" transition and it will turn out that revealing the true character of the two interactions will be of great importance to our knowledge of the neutrino.

Majorana elaborated a symmetrical theory of weak interaction in 1937, where the genuinely neutral neutrino, with no quantum numbers to conjugate, is indistinguishable from its anti-particle. This possibility is thence named the "Majorana neutrino". It is still not clear whether the SM is right in preferring the "Dirac neutrino", which is genuinely different from its anti-particle, to the Majorana neutrino. We shall pay special attention to this question when addressing the issue of the neutrinoless double beta decay.

In 1937 muon has been discovered in cosmic rays and this opened route to the second neutrino flavor (expressing everybody's consternation with this discovery Rabi is told to have exclaimed in dismay: "Who ordered that?!"). In 1941 it was shown to decay into one single electron and, since nothing else could be seen, the physicists, already accustomed to the idea of a neutrino, suspected that another decay to yield a neutrino has been found (until then the nuclear beta decays were the only known source of neutrinos). When in 1948 the electron spectrum from muon decay was found to be continuous it became obvious that not one but two neutrinos are emitted along with the electron. Pontecorvo witnesses that at that time everybody felt that the two neutrinos should be different. They were even named differently, the "neutrino" and the "neutretto", but with time the idea seem to have been forgotten and it was only in 1962, when the difference between the two neutrinos has been clearly demonstrated in the first of a long series of important accelerator neutrino experiments, that the electron and the muon neutrino were finally given life.

1938 saw the advent of the Bethe theory of stellar thermonuclear synthesis, necessarily based on the weak reaction:

$$p \rightarrow n + e^{+} + \nu, \qquad (4)$$

which is mass forbidden for free protons but may go if the neutron is simultaneously captured into a nucleus (another proton), when the emitted binding energy makes up for the missing mass. This reaction is supposed to be the main source of solar neutrinos, later to play crucial role in the development of neutrino astrophysics, and also the strongest source of low energy neutrinos on Earth, since the positron beta decay and electron capture nuclei (which do not exist in natural radioactivity) are not easy to produce in great abundance. If we take that the solar luminosity, which yields about 1 kW per m$^2$ on Earth, is ultimately due to the process of hydrogen burning in which four protons turn into one alpha particle, two positrons, two neutrinos, and 26.7 MeV, then it is straightforward to find that the solar neutrino flux on the surface of the Earth should be about 6.5x10$^{10}$ neutrinos per cm$^2$ per second. Real situation is much more complex and the attempts to measure this flux, some thirty years later, produced the famous solar neutrino problem, which is eventually nearing solution only today.

Another event which is today still of interest to the story of the neutrino is the first discussion of the neutrinoless double beta decay by Furry in 1939, after Goeppert-Mayer, at the suggestion by Wigner, discussed the ordinary two-neutrino double beta decay in 1935. We shall dwell on the details of this process when discussing the issue of neutrino mass.

In 1942 Fermi invented and constructed a nuclear reactor, the most powerful source of low energy electron antineutrinos on Earth (apart from a nuclear explosion), and thus made another great contribution to neutrino physics. It was learned that reactors deliver about 2x10$^{17}$ of roughly MeV (anti)neutrinos per second per 1 MW, what enabled the first direct observation of the (anti)neutrino, some ten years later.

Then there followed a long series of difficult experiments on nuclear recoil measurements in beta decay, which first served to further strengthen the indirect evidence of the existence of the neutrino, and then to establish the form of the angular correlation between the beta electron and the neutrino. This correlation, of the notorious $1+A\cos\theta$ type, is among the few observables sensitive to the actual form of the weak interaction Hamiltonian, or to the behavior of the weak currents under Lorentz transformations and coordinate inversion. The Fermi current has to be of either of the V(ector) type, as Fermi supposed originally, or of the S(calar) type, while the GT current has to be either of the A(xial vector) type or of the T(ensor) type. The weak Lagrangian, which contains the product of the total current with itself, was thought to has to be a scalar, in order not to change sign upon the transformation of r into −r, and thus to describe the parity conserving interaction. The recoil experiments, though often yielding inconclusive evidence, were found to support the S and T character of the interaction, thus corroborating the required scalar nature of the Lagrangian (the issue of the so called Fierz interference between the V and S, and A and T terms, which was later shown to be absent, only complicated the matters). This well known episode, which may suitably be termed as "The Comedy of Errors", lasted for some ten years, all the way to 1956, to be settled only by the overthrowing of the conservation of parity in weak interactions.

State-of-the-art in the field in 1949 resulted in the formulation of the universal Fermi interaction, the herald of the weak interaction (Fig.1), when the nuclear beta decay(s), the muon decay and the capture of the muon by a nucleus were all recognized to be varieties of the same interaction.

In 1952 the half-life of the beta decay of the free neutron was measured, what enabled reliable estimate of the cross section for the capture of the (anti)neutrino by a proton, reaction (3). By predicting the rate for this reaction this opened the way to the (anti)neutrino detection. On the experimental side it was enabled by the development of scintillation detectors and of high power reactors.

In 1953 Marx, Zeldovich, and Konopinski and Mahmoud, formulated the law of lepton number conservation (the electron lepton number of today), which for the first time introduced a clear-cut definition of the neutrino and the antineutrino (though Lee witnesses that Fermi had an idea of both the baryon and lepton number conservation already in 1948). The law ascertains that in the $\beta^-$ decay the antineutrino is emitted by the neutron and that in the inverse reaction it may be captured only by the proton and not by the neutron.

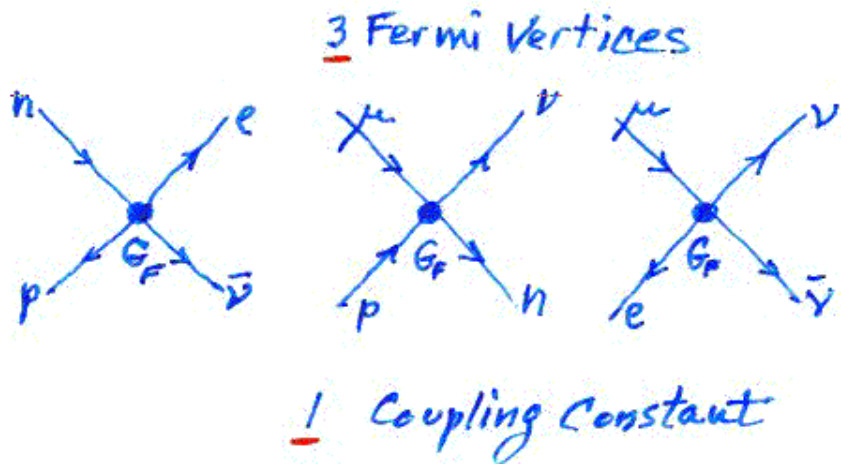

**Fig.1.** Recent hand-drawing by Jack Steinberger [5], illustrating the universal Fermi interaction of 1949 in the language of Feynman diagrams, which were devised at about the same time.

All this finally led to direct detection of the reactor antineutrino via the capture reaction by a proton [Eq.(3)] in the famous experiment performed by Reines and Cowan in 1955 (what turned out to be worth the Nobel prize only 40 years later - the second Nobel prize for a neutrino!). The two annihilation quanta followed after thermalization time by neutron capture gamma rays provided the unambiguous signature which was occurring at the theoretically predicted rate. The experiment opened the era of large, high efficiency, electronic detectors in particle physics. When in the same year Davis failed to detect the same antineutrinos by the radiochemical method suggested by Pontecorvo in 1946, using the capture reaction:

$$\bar{\nu} + {}^{37}\text{Cl (n)} \rightarrow {}^{37}\text{Ar (p)} + e^- \qquad (5)$$

the neutrino and antineutrino were thought to have been proved genuinely different particles (Dirac particle). At the same time this proved the conservation of the lepton number. Everything seemed neat and the chapter of the neutrino seemed practically closed. Nobody suspected the stormy events which were brewing around the corner, in the quickly developing field of elementary particle physics.

A.1.2. Second epoch: 1956-1958
> *Weak interactions are found not to conserve parity, the two-component theory of the neutrino is justified and V-A character of weak interactions established*

The unsuspected crucial change in our understanding of the weak interaction and of the neutrino, which up to this moment came mostly from the study of nuclear beta decays, was induced by problems encountered in the study of elementary particles. The two strange particles named the θ and the τ, which judging by their (slow) decay modes possessed different intrinsic parities, were after detailed study found to be by all other characteristics, in particular the lifetime and mass, absolutely identical. This was known as the θ-τ puzzle. In 1956 Lee and Young noticed that in the processes governed by the weak interaction the conservation of parity has never been explicitly tested and suggested that the θ and the τ could possibly be one and the same particle which decays via the interaction which does not conserve parity. Even before parity conservation in other weak processes was purposefully experimentally tested Lee and Young, Landau, and Salam developed the so-called two-component theory of the neutrino, which did not conserve parity. It followed the theory developed by Weyl back in 1929, when it was discarded because of the asymmetry it introduces, but which now seemed appropriate. Instead of the four states of the Dirac theory describing the neutrino and antineutrino of both helicities, the two-component theory considers only the massless neutrino of one helicity and the antineutrino of the opposite helicity as the physical states. Parity operation on those states would produce the states which do not occur in nature, what violates the invariance under coordinate inversion. This turned out to be the description of neutrinos which is still valid in the SM.

Then, in 1956-8, came a series of ingeniously designed experiments to confirm the conjecture that parity is not conserved in all weak processes. They managed to measure the pseudoscalar quantities which were never measured in the weak processes before. The pseudoscalar, i.e. the scalar product of an axial and one polar vector (as is the scalar product of spin and momentum, like angular distribution or helicity) changes sign upon parity operation, and, if the interaction is to conserve parity, the pseudoscalars must average to zero. The Wu experiment measured the asymmetry of the angular distribution of electrons emitted from the ensemble of oriented beta decaying nuclei (an unsuspected effect in the decay of probably the most widely used of all the radioactive isotopes!), the Goldhaber experiment measured always one and the same helicity of the neutrino emitted in beta decay (as transferred to the subsequently emitted gamma-ray, in the only favorable case for this kind of measurement offered by hundreds of beta decaying nuclei!), and a number of experiments tested parity conservation in weak particle decays, notably the Λ and the π (the θ and the τ, whence it all started, became a single K). All the experiments tested the processes which involve the charged weak currents (the only known at that time) and all of them found maximum possible violation of parity conservation.

Main results of interest to our story from this campaign were that the (electron) neutrinos emitted in all the charge changing weak processes are invariably of negative (or left-handed) helicity, while the antineutrinos are right-handed, of positive helicity. The two-component neutrino theory was thus justified and corresponding modifications in the theory of weak interactions were required. The repeated nuclear recoil correlation experiments this time confirmed that the weak interaction currents are indeed of the Vector and Axial vector type (and not S and T, as was found before) and time was ripe for what has since become known as the V - A theory of weak interactions. It was formulated in 1958 by Feynman and Gell-Mann, and Marshak and Sudarshan.

The variant of the theory was chosen which defines the weak current as the difference of the vector and axial vector currents taken with equal weights. The so-called current-current hypothesis interpreted all the weak processes as arising from the interaction of the total weak current of the V - A structure with itself. This form of the current describes the coupling to left-hand particles and right-hand antiparticles, while the sum of the currents would describe the opposite case. Equal weights correspond to maximum parity violation since this leads to maximum values of mixed terms in the product of the current with itself in the weak Lagrangian, which change sign upon coordinate inversion. (It is important for us that for neutrinos even the neutral currents have the same structure - neutral currents of charged leptons and quarks are different linear combinations). It described well all the weak processes known at that time and this closed another chapter in the story of the neutrino. However, like the original Fermi theory, it was not renormalizable, leading to divergencies at high energies. Subsequent theoretical development had to take care of that, what, combined with the development of many powerful experimental techniques, led to significant new knowledge of the neutrino and its place in the workings of nature.

Before we proceed further two comments are perhaps appropriate here. The first concerns the issue of the difference between the neutrino and its antiparticle. We saw that the Davis' unsuccessful attempt to detect the (reactor) antineutrino in 1955, by the capture by the neutron [Eq.(5)], demonstrated the genuine difference between the neutrino and the antineutrino. However, the finding that in the weak interaction only the left-handed particles and right-handed antiparticles participate opened for massless neutrino (what is the necessary condition for its helicity to be a constant of motion) the possibility to be identical to its antiparticle (to be a Majorana and not a Dirac or, more precisely, a Weyl particle) the difference in their interaction ability then being only due to the so-called "helicity mismatch" in the emission and absorption vertices (the lepton number is even then convenient to introduce to distinguish between the helicity states, and of course as the flavor identifier). Finite neutrino mass, on the other hand, would introduce into the propagating neutrino the admixture of the opposite helicity state, proportional to 1-v/c (or to the "chirality factor" $(m_\nu/E_\nu)^2$), due to the possibility for a particle to overtake the neutrino and see its helicity reversed. For a Dirac neutrino this would be the state considered to be unphysical, and thus unobservable, or "sterile", as it is nowadays called, while for a Majorana neutrino this is a perfectly regular neutrino state, hard to observe only due to the smallness of its amplitude (due to the smallness of the neutrino mass the admixture of the "alien helicity" is very small at all but the lowest energies, when on the other hand the cross sections are small, so that it is everywhere similarly hard to observe). Matters are additionally complicated by the possibility for the weak interaction, as suggested by some GUTs, to have a right-handed component (and thus again hardly observable at low energies). To distinguish between the possibilities is so difficult that we still do not know which is the right one. We shall consider this problem in some more detail again when dealing with the case of the neutrinoless double beta decay.

The second comment, which is related to the first one, concerns the violation of discrete symmetries, other than parity, by weak interaction, and by neutrino behavior in particular. It is obvious that charge conjugation, which changes the (Dirac) neutrino into an antineutrino and vice versa, but does not influence its helicity, produces an unphysical state in the same way as does the parity operation. Both P and C invariance are thus violated for the neutrino, while the CP is still thought to be conserved, though, minding the smallness of CP non-conservation for the K, it is easy to overlook any similar effect in neutrino interactions, where the statistical accuracy of that order is extremely difficult, if not impossible, to attain. Like other ambiguities about the neutrino of similar magnitude this one survives to this day from the period we have just reviewed. Similar problem arises if the neutrino has a finite mass and, as discussed above, possesses an admixture of opposite parity. This would be equivalent to incomplete violation of P or C conservation. Thus, if any effect is found which this time violates (only slightly) the behavior of the neutrino as described by the V - A interaction, it would be difficult to discriminate between a number of possible causes. As we shall see, similar problems only multiply in possible extensions of the SM.

A.1.3. Third epoch: 1958-1983
*Muon neutrino is proved and tau neutrino suspected. Electroweak interaction is contemplated and fully established. Neutrinos start doing some useful work.*

First great event in neutrino physics after the "parity furore" of 1956-58, was the establishment in 1962 of the second neutrino, the $\nu_\mu$, upon the suspected existence of which in the late forties we have already commented. In the meantime major experimental developments took place, which included the production of (muon) neutrino beams at high energy proton accelerators by the copious pionization on thick targets and subsequent pion decay in flight into a muon and a beamed (muon) neutrino, as well as further detector developments, the most illustrious in that period being the big real time track detectors, the spark and the bubble chamber. This marked the beginning of accelerator (muon)neutrino experiments which contributed so much to both the weak and strong interaction physics in the years to come. Lee and Young gave another stimulus to experimental enterprises by calculating the cross sections for accelerator neutrino induced reactions. Those were in the first place the high energy neutrino "capture" reactions by nucleons of the type

$$\nu_l + n \rightarrow p + l^- \qquad (6)$$

where $l^-$ would stand for either electron or a (negative)muon, with the estimated cross section of some $10^{-38}$ cm$^2$ at 1 GeV. This reaction was to decide whether the two neutrinos (actually the neutrino and the antineutrino) emitted in the muon decay are identical or not (of the same flavor, in modern terminology). The muon was being thought to have the same lepton number as the electron, and was thus considered merely a "heavy electron" (or, better still, an excited electron), which should, if the two neutrinos were identical, as well decay into an electron and a photon. The absence of this decay mode strongly suggested that the two neutrinos are genuinely different particles, and that the muon is not just a heavy electron but is a particle of its own (as the emission of a photon, which is identical to a particle-antiparticle pair, can not change the nature of the particle). If the neutrino which accompanies the muon in both the muon and pion decay is indeed different from the one which accompanies the electron in beta decay it would in reaction (6) produce only muons and no electrons (as suggested by Pontecorvo in 1959). This is exactly what Lederman, Schwartz and Steinberger found in 1962, on the basis of the sample of 40 recorded neutrino induced events (first Nobel prize for a neutrino, 1988). The second generation of leptons was thus completed (though not yet named that way), and the need for two lepton numbers was demonstrated, one for each family, which would most probably be independently conserved.

We now devote some attention to events from hadron physics of that time which are seemingly not on the main line of our story but which turned out to be of great importance for the development of concepts which later led to great unification of particle physics. In 1963, to save the universality of the weak interaction and account for the extremely slow decays of strange particles, Cabibbo introduced the fruitful concept of what is nowadays called the quark flavor mixing by the weak interaction. Weak interaction was found responsible for the long lifetimes of strange particle semileptonic decay modes (in which one vertex is hadronic and the other leptonic), the hyperon decays being much slower than predicted by the universal Fermi interaction. Cabibbo supposed the charged weak hadronic current part of the total current to be composed of both the strangeness conserving and the strangeness changing terms, the weights of which were expressed through the value of the "Cabibbo angle", as determined to fit reality (if the strange particles were stable the Cabibbo angle would be zero, or the strangeness changing processes would be totally suppressed). It turned out that a unique value of this single parameter (and the ratio of axial to vector coupling constants) reproduced well the lifetimes of not only strange particle semileptonic decays, but the lifetime of the neutron as well, saving at the same time the universality of the weak interaction. In terms of the quark structure of hadrons this was later formulated as quark flavor mixing by the weak interaction, the $s$ quark being mixed with the $d$ quark to the extent determined by the Cabibbo angle, into the "rotated" $d'$ and $s'$ weak interaction eigenstates of indefinite mass. The scheme, however, predicted the existence of weak neutral strangeness changing current, which could not be observed ($d \leftrightarrow s$ transitions here, but more generally - in contrast to the charged current interactions, no neutral flavor changing currents exist). The problem was solved in 1970 (after the advent of electroweak theory) by Glashow, Illiopoulos and Maiani, who introduced the fourth quark, the "charm", or the $c$, to complete the second quark doublet (charm has been suggested to exist by Glashow and Bjorken as early as 1964, for the number of weak quark currents to get equal to that of lepton currents). The neutral currents constructed from the two doublets then cancel exactly (the "GIM mechanism"), the main weak transition of the $c$ being that into the $s$ (though, again, the flavor changing neutral currents are predicted by some GUTs and small admixtures could perhaps be observed at low energies). The idea of quark-lepton symmetry was thus established, what clearly demonstrated for the first time the unity of all the particle phenomena. The formalism has been extended by Kobayashi and Maskawa to the case of three families in 1973 [Eq.(2)], though the third family was clearly suspected to exist only in 1975, upon the unexpected discovery of the heavy $\tau$ lepton (for the b quark was not found before 1977). Another, even more explicit link with neutrino physics emerged when the neutrino scattering experiments turned out to provide first indirect experimental evidence of charmed baryons. Namely, the charmed particles (like the strange ones) must in strong interactions, which conserve both the c and s, be produced in pairs (the historical "associated production") but in weak interactions, which change both the strangeness and charm, the neutrino may produce a single charm, thus halving the needed energy. It was in muon neutrino scattering on nuclei that the products of the interaction were interpreted as the decay products of the new charmed particles. Those were among the first in the row of neutrino scattering experiments which contributed to hadron physics - neutrino started to do some useful work!

1967 saw the major theoretical breakthrough in the form of the Glashow-Weinberg-Salam unified electroweak interaction. From the formal viewpoint it is the (renormalizable) local non-Abelian gauge theory of the $SU(2) \times U(1)$ symmetry, with a massless boson singlet and a triplet, the singlet remaining the massless photon $h\nu$, while the triplet ($W^+$, $Z^o$, $W^-$) acquires mass through the supposed Higgs mechanism of spontaneous symmetry breaking. Proper mixing of neutral bosons, adjusted to fit the experiments (by fitting the weak mixing, or Weinberg, angle, $\theta_W$), gives the physical neutral bosons, the $h\nu$ and $Z^o$. Differences from previous theories in the weak sector manifest mainly at energies higher than the boson masses, above some 100 GeV, as well as in the existence of weak neutral currents (which were, to do justice, suspected to exist much earlier). In the low energy limit, when the momentum transfer is much smaller than the boson mass, the theory reduces to the Fermi four point interaction, or to the V - A variant of that theory (Fig.2). This is where we learned that the weakness of the weak interaction, and of the neutrino interactions in that number, is not due to the smallness of the coupling constants but rather due to the enormous masses of the intermediate bosons and the correspondingly small range of the interaction. After the Fermi theory of beta decay and the non-conservation of parity this was no doubt the most important single achievement in our understanding of the weak interaction in general, and of the neutrino in particular. It was to be experimentally confirmed in the period from 1973, when the neutral current neutrino interactions were detected, to 1983 when the real intermediate bosons were observed. The Higgs boson, however, which is essential to the theory, still awaits detection (mind, though, the "closing down of the LEP affair", when LEP has been closed in the year 2000, to give way to LHC, in spite of the claims that the decay of the Higgs has been seen at about 115 GeV, but at the level of 2.9$\sigma$ above background instead of at the (arbitrarily) required level of 5$\sigma$, and the pleads to continue the measurements to improve on the statistics).

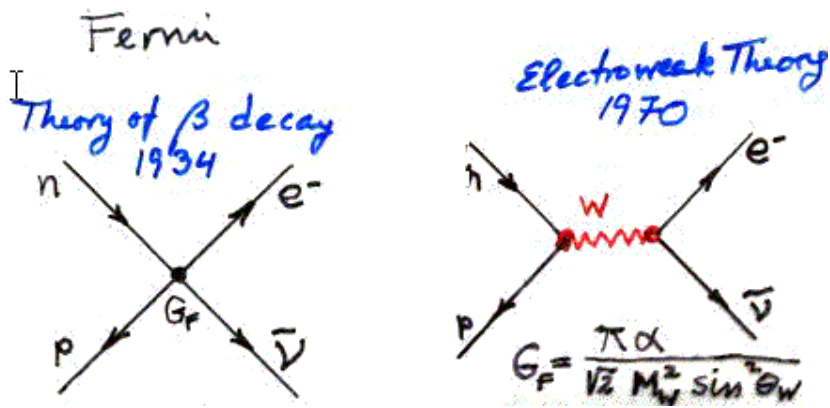

**Fig.2.** The difference in the views of the Fermi weak interaction theory and electroweak theory of Glashow, Weinberg and Salam, illustrated by another of Jack Steinberger's hand-drawings.

It was only in 1969 that the electron neutrino was conclusively detected for the first time, and that necessarily had to be the solar neutrino! With all his experience with the Chlorine-37 radiochemical method [Eq.(5)] for the detection of the reactor antineutrino, which he did not see, Davis boldly set to measure by the same method the flux of neutrinos coming from the assumed thermonuclear reactions running in the Sun's interior. The threshold for the detection reaction [Eq.("anti5")] is 0.8 MeV, which is above the continuous spectrum of neutrinos from the basic pp reaction, what makes it sensitive to neutrinos from the decays of $^7$Be and $^8$B only (see Fig.3, where the neutrino fluxes as calculated within the rather stringent limits of the Standard Solar Model (the SSM) are presented). The predicted reaction rates vary from 6 to 8 SNU (1 Solar Neutrino Unit (SNU) = 1 event per $10^{36}$ target atoms per second), while the average measured rate, after some 30 years of measurement (!), equals 2.5 SNU, with an error of 10%. This constitutes the famous solar neutrino problem. The Davis' chlorine experiment was unique in many respects - it was the first to bring the enormous detector deep underground to reduce the background induced by cosmic-ray muons, and, though constantly improved, it may be looked upon as the longest lasting single complex measurement in the history of mankind. It marked the beginning of the new field of physics, or of the neutrino physics - the neutrino astrophysics. Many followed and the field grew rapidly, yielding the results which we shall necessarily comment later on.

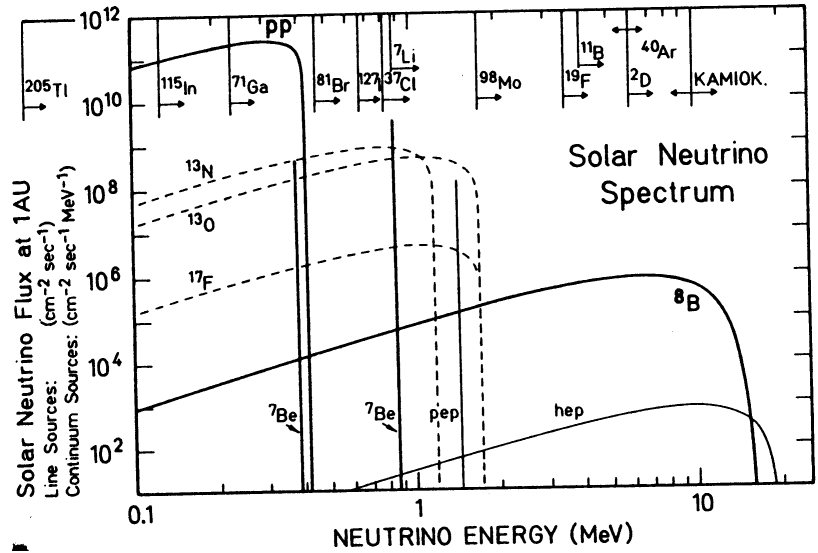

**Fig.3.** The spectrum of solar neutrinos as calculated within the Standard Solar Model for the basic pp chain which by assumption produces more than 98 % of solar energy [23]. Thresholds for different solar neutrino detectors, on which we shall comment later on, are also marked.

In 1973 the muon neutrino reactions with no charge change were observed for the first time in the monstrous "Gargamelle" bubble chamber. These "non-muonic" muon neutrino reactions, in which muons are not produced, of the type

$$\nu_\mu + e^- \to \nu_\mu + e^- \qquad (7)$$

(elastic scattering, with cross section of the order of $\sigma/E_\nu \approx 10^{-42}$ cm$^2$/GeV) and

$$\nu_\mu + p \to \nu_\mu + p + \pi^+ + \pi^- \qquad (8)$$

(deep inelastic scattering, with the cross section of the order of $\sigma/E_\nu \approx 10^{-39}$ cm$^2$/GeV) were the only possibility to observe the weak neutral currents which are in other cases, which involve charged particles, hidden behind the very much stronger effects of the electromagnetic interaction. They are supposed to be mediated by the $Z^o$, the weak neutral intermediate boson of the electroweak theory, whose mass in relation to the mass of the charged W boson and whose coupling to the weak neutral currents in relation to the charged current coupling is determined by the value of the Weinberg mixing angle. The measured cross sections enabled the single value of the mixing parameter to be determined, which turned out to satisfy the whole universe of the weak interaction data (though some recent high precission data suggest that the Weinberg angle as determined from the neutrino scattering might be slightly but significantly different from the values extracted from other measurements [40]). The electroweak theory was fully established and it only remained to create the free intermediate bosons at the masses predicted by the value of the mixing angle.

During the next couple of years experimental study of muon neutrino reactions contributed greatly to hadron physics. In the first place the finding that the cross-section for neutrino inelastic scattering on the nucleon increases linearly with energy confirmed the point-like parton or quark structure of a nucleon. Then in 1975 comparison of electron and neutrino scattering on nucleons, as well as the study of structure functions in neutrino and antineutrino scattering on nucleons, yielded quark electric charges; experiments which perhaps came nearest to direct observation of quarks. The already mentioned production of charm in neutrino reactions comes into this category.

In this era of eruptive proliferation of elementary particles the search for a lepton heavier than the muon was only to be expected, though nothing in theory pointed to its existence. Such a particle was systematically searched for among the decay products of heavier particles as well as in known lepton production interactions, only at higher energies. Until 1975 it was not found up to about 1 GeV what, assuming weak interaction universality, by direct scaling of the muon half-life for the phase space factor, $(m_x/m_\mu)^5$, left no hope that it could be directly observed. The search in high energy electron-positron collisions, in which creation of the lepton-antilepton pair was expected, was thus based on the search for its assumed decay products, the lighter leptons (though at those energies decay into hadrons, a "semihadronic" decay, was also expected). The MARK I detector at Stanford, one of the first big complex electronic detectors, which already did good job observing charmonium, was engaged by Perl and his team for this search. At the threshold energy of 3.56 GeV the simultaneous occurrence of an electron and a muon started signaling the creation of a tau-antitau pair (the τ, for the "third", in Greek). At 1.78 GeV the heavy lepton turned out heavy indeed. This discovery immediately implied the existence of the third generation of fundamental particles, of the corresponding neutrino in the first place, as well as of the third quark doublet, the completion of which took another 25 years of hard work, the tau neutrino being observed last, in the year 2000. As it will turn out, we have reasons to believe that there are no leptons heavier than the tau.

Finally, in 1983 the decays of the free intermediate vector bosons of the weak interaction were observed at the purposefully constructed marvel of accelerator technology, the $p\bar{p}$ collider at CERN, by the UA1 and UA2 collaborations concentrated around the sophisticated detectors of the same name. With the estimated lifetimes of the order of $10^{-24}$ s the bosons themselves are impossible to observe even when created free but their decay channels into leptons leave clear enough signatures amidst the chaos of hadron events. The observed decay of the W into an electron (and a neutrino) and that of a Z into an electron or muon pair (five of them observed altogether!) enabled the reconstruction of the invariant boson masses which turned out to agree well with those expected from the electroweak theory and the known value of the Weinberg angle. This earned the next year's Nobel prize to Rubbia and van der Meer. In the years to come our knowledge of the gauge bosons increased, especially with the introduction of LEP, when the masses and widths of the bosons could be measured with high precision. This enabled the precise evaluation of contributions from higher order diagrams to the weak coupling constant, in particular of those involving the yet unobserved t quark, what led to reliable estimate of its mass. As far as our story of the neutrino is concerned precise measurement of the shape of the $Z^o$ resonance brought about the most important conclusion: this shape, or the probability for the $Z^o$ to disappear without trace, turned out consistent with its decay, among other modes, into exactly three light neutrino flavors, the decay into two or four neutrino pairs being well outside the limits of error. This left in the period to come only the tau neutrino to be found. However, the true nature of the neutrino is still not at all certain! To this we devote our next chapter.

A.1.4. Fourth epoch: 1983-2001
> *Tau neutrino is observed and the three generations of fundamental particles completed. The limits of neutrino parameters are narrowing, but,...the true nature of the neutrino remains uncertain.... Neutrinos continue to do useful work.*

Nearing the present day the events in the story of the neutrino multiply and it is becoming increasingly difficult to discuss them in chronological order. In this chapter we shall thus first review the completion of the third generation of fundamental fermions and then comment upon the more recent development of our knowledge of the detailed properties of the neutrino, as well as on the most important neutrino properties relevant to other fields of knowledge.

Meeting the expectations, the t quark has been observed at Fermilab in 1995, thus completing the third quark doublet and strengthening the belief in the existence of the tau neutrino. According to the SM the tau neutrino would be the last missing fundamental fermion - it would at the same time complete the third lepton doublet and possibly close the list of truly elementary particles. During the five months campaign in 1997 the "Direct Observation of NU Tau" collaboration was exposing the sophisticated nuclear emulsion target to the intense beam of neutrinos at the 800 GeV proton beam at the Tevatron. The target was a stack of sandwiches of thin emulsion layers separated by millimeter steel plates to serve as the interaction material, all this interleaved with scintillation fiber tracking planes to indicate and record the coordinates of the charged particle tracks. The downstream spectrometer complex provided particle identification. The neutrino beam by assumption contained some 10% of tau neutrinos, originating mostly from the decay of the heavy $D_s$ ($c\bar{s}$) meson. The tau neutrino would upon interaction of the type (6) in the steel plate hopefully produce a tauon. At its short lifetime the tauon would at most cover a couple of millimeters before decaying into either an electron, a muon, or a hadron, and the neutrinos. The tauon which would reach the emulsion would leave a short track extending into a long track after a characteristic kink. Discrimination against the background events due to neutrinos of other flavor was performed according to the magnitude of the transverse momentum imparted to the kink forming particle. Following a long search by automated microscopes of the regions pinpointed by the tracking detectors four events were found which stood all the tests and led to the announcement of the positive result in July 2000. Besides, the experiment declared the comeback of the nuclear emulsion in the new guise, some fifty years after the glorious service in the period of pre-accelerator cosmic-ray particle physics.

The one hundred years long quest for the fundamental building blocks of nature thus seems to has come to an end (mind the Higgs, however!). To our satisfaction the book, which was opened with the discovery of the electron, has been closed by the establishment of the one last type of the neutrino. Whether the book is indeed closed not to be opened again, or not, is not for us here to discuss.

a. The properties of the neutrino

In what follows we shall merely try to review the still open questions pertaining to the physics of the neutrino. These questions concern the detailed properties of the neutrinos: their masses and the related question of their stability, their electromagnetic moments, and their "Diracness", as well as the possible and yet unknown peculiarities of the interactions by which they so scarcely communicate with the rest of the world. After this we shall commit ourselves to the shortest possible review of that part of the new field of astroparticle physics which is connected with neutrino physics.

a.1. Neutrino masses

We start with the far reaching problem of neutrino masses. To get the impression of the complexity of the problem it is perhaps good to examine the chapter on the neutrino in the full Review of Particle Physics by the Particle Data Group (PDG), 2000 [13]. Of all that richness we shall necessarily comment on only what may be considered essential. As we already said, the masses of neutrinos of all flavors are in the SM assumed to be exactly null. This, however, is not supported by any local symmetry of SM theories and some GUTs give finite masses to neutrinos. Any sign of non-zero neutrino masses thus signalizes new physics beyond the SM. Be as it may, the issue is to be settled only experimentally and ever since the invention of the electron neutrino attempts to determine its mass constituted an important activity in neutrino physics. To grasp at least some of the subtleties of this complex problem we have to deal with possible definitions of neutrino masses in some detail. Namely, in addition to conventional possibility for a physical particle to have a unique rest mass, what for a long time went without saying for the neutrinos, if they posses mass there is room left for neutrino flavor mixing by the weak interaction, what would, minding the analogy with the quark sector, make the quark-lepton symmetry more complete. This would imply the existence of the basis neutrino states of definite mass (the mass eigenstates) in which neutrinos propagate in empty space, usually denoted as $\nu_1$, $\nu_2$ and $\nu_3$, with masses $m_1$, $m_2$ and $m_3$, which would be mixed (or rotated) by the weak interaction into the weak interaction eigenstates of indefinite mass, which participate in the weak vertices, and which we denote as our familiar and observable $\nu_e$, $\nu_\mu$ and $\nu_\tau$. The extent of the mixing would be described by the matrix analogous to the CKM matrix (2), sometimes abbreviated MNS, after Maki, Nakagawa, and Sakata, who were the first to write it down as early as 1962 [14]. In the most often considered case of the mixing of two lightest neutrinos of masses $m_1$ and $m_2$ into the electron and muon neutrino the magnitude of the mixing is described by the single mixing angle $\theta$, analogous to the Cabibbo angle:

$$\begin{aligned}|\nu_e\rangle &= \cos\theta\ |\nu_1\rangle + e^{i\rho}\sin\theta\ |\nu_2\rangle \\ |\nu_\mu\rangle &= -e^{-i\rho}\sin\theta\ |\nu_1\rangle + \cos\theta\ |\nu_2\rangle\end{aligned} \qquad (9)$$

where the phase factor $e^{i\rho}$ can equal only 1 or $i$, if CP invariance is assumed. This scheme offers a number of interesting consequences which we shall now discuss.

Neutrino masses would manifest themselves in an observable way in variety of phenomena. The weak nuclear and particle decays, in which the energy and momentum of observable decay products depend on the emitted neutrino mass, constitute the first class of such processes. The methods which exploit this exactness of uncorrupted kinematics are called "direct" though, clearly, do not observe the neutrino at all. The mass of the electron neutrino is best determined by the Fermi-Kurie analysis of allowed beta spectra, while those of the muon and tau neutrinos follow from the analysis of the decays of kaons, pions and tauons. All these measurements up to this date produced only upper limits for the masses, the best values of which are: 2.5 eV, 190 keV and 18 MeV for the e, mu and tau nu respectively. Those figures certainly call for a number of comments.

There is also a number of processes which would not be taking place at all if the neutrinos were massless. Mere observation of these processes would testify of finite neutrino masses while their amplitudes would yield the masses themselves. Methods based on the study of these processes are termed "indirect". The neutrinoless double beta decay, neutrino oscillations, neutrino decays and electromagnetic interactions of neutrinos fall into this category. Up to now no consistent conclusions about the masses have been reached from the study of these phenomena, though some recent results seem to more significantly support their non-zero values. We shall also briefly comment on the most important of these results.

a.1.1. Direct methods to measure neutrino masses

Let us first comment on the results of direct measurements of neutrino masses - of the electron neutrino mass in particular, which are statistically most significant and are, since are related to the first and possibly only stable lepton family, also the most interesting ones [9]. That the measurements involving neutrinos are extremely difficult and prone to misinterpretations we have already seen, and will still see, but direct measurements of their masses, due to their obvious smallness, are perhaps the most so. Best results for the electron neutrino have so far been produced by meticulous measurements of the tritium beta spectrum. Assuming zero mass for the neutrino the Fermi-Kurie plot for allowed decays should be a straight line all the way down to its intersection with the energy axis at the energy available to the decay. For the non-zero unique neutrino mass (without mixing) the spectrum falls short to this point by the value of the mass, and thus departs from the straight line as it nears its end. Finite energy resolution, clearly, introduces opposite curvature into the end of the spectrum, making the deconvolution of the transmission function only the first among the corrections essential to recovery of the true spectrum. For the two neutrino mixing case [Eq.(9)] the term responsible for the departure of the plot from linearity is of the form:

$$\cos^2\theta \,[\, \Delta^2 - m_1^2 \,]^{1/2} + \sin^2\theta \,[\, \Delta^2 - m_2^2 \,]^{1/2} \qquad (10)$$

where $\Delta = E_o - E_e$ and $E_o$ is the energy available to the decay. In the case of no mixing ($\theta = 0$) this reduces to $[\Delta^2 - m_{ev}^2]^{1/2}$, as described above. Recovering the shape of the spectrum thus yields the neutrino mass(es) squared, which may come out even negative, if the endpoint overshoots the decay energy. Needles to say that this is what actually happens. The real mass solution is then obtained only as an upper limit of the confidence interval which at a given confidence level encompasses zero, and this is what is conventionally cited as the result from such measurements. After a number of exciting episodes and turnovers the world average for the square of the mass of a unique (unmixed) electron neutrino stabilized at the 68%CL at something like $-3.5(57)$ eV$^2$/c$^2$. This then yields the cited result $m_{ev} < 2.5$ eV/c$^2$ at the 95%CL. Though this result is the most reliable yet, it is not at all free of ambiguities. In particular, there is a certain structure in most of the data near the endpoint which bears the name of "anomalies", which stubbornly remains there after all possible corrections are made, and is greatly responsible for the negative square of the mass. Whether it is due to some artefact yet unaccounted for or reflects relevant physics is not clear. The extreme view, which we quote to illustrate the wealth of possible solutions, is to interpret this feature as being due to the tachyon nature of the neutrino, in which case its mass would be imaginary and mass squared negative. Rather consistent scenarios have been contemplated to accommodate such a superluminal neutrino (including the unexpected explanation of the origin of the depletion of the primary cosmic-ray proton spectrum above some $10^{17}$ eV (the so called "knee") as being due to the decay of such ultra-high energy free protons [Eq.(4)], which would become possible in reference systems in which tachyon neutrino energy becomes sufficiently negative [10,11]). From the trend of the results, however, it is rather obvious that we are still far from reaching the interval which would not embrace zero at a satisfactorily high confidence level. Whether we shall ever reach such an interval is actually the right question to ask, and it is the answer to this question which will greatly determine the future of neutrino physics. If the neutrinos were indeed massless, as the SM assumes, or extremely light, as some GUTs predict, the physics of the neutrino would be in the real danger of getting entangled in the eternal asymptotic pursue of zero or near zero values. Luckily though, as we shall see, the indirect methods may supply a definite answer sooner. Of other curiosities which emerged from direct measurements we mention the at one time seen structure at the very beginning of the tritium spectrum which has been interpreted in the two flavor mixing scheme as being due to the heavy neutrino, a possible main component of a muon neutrino, admixed with couple percent to a neutrino of vanishing mass, which would again be the main component of the electron neutrino. This "17 keV neutrino" episode seem to have been resolved by dismissing this possibility, the effect most likely being due to electron scattering on entrance diaphragms. Direct measurements of muon and tau neutrino masses are of even much lower sensitivity and at present render high and almost useless upper limits which we shall not discuss further.

a.1.2. Indirect methods to measure neutrino masses

We now move on to explore in some detail the equally slippery ground of indirect neutrino mass measurements. In this we shall necessarily touch upon the problems concerning neutrino properties other than mass and eventually get the idea of full complexity of the frequently all too easily and loosely used term - the "neutrino".

a.1.2.1. The neutrinoless double beta decay

We first discuss the problem of the neutrinoless double beta decay [37]. The double beta decay (the "$2\nu 2\beta$" decay), which is a second order process where two nucleons (neutrons or protons) simultaneously emit a real lepton pair each, has by now been observed in a number of favorable cases, including those where this is neither the only decay mode nor the fastest one, like U-238. Minding the negligible decay rates on the experimental side and problems with nuclear matrix elements on the theoretical side, measured lifetimes, which are in the range of $10^{19}$-$10^{21}$ years, agree satisfactorily with theory and the process is apparently understood well. In the supposed neutrinoless variety (the "$0\nu 2\beta$" decay) only real electrons are created while the two nucleons exchange a virtual neutrino. The electrons would thus carry the decay energy, what is a nice signature and makes up for the low probability of the process, which, however, may eventually reach the probability for the $2\nu$ decay due to having only three instead of five particles in the final state. The problem is, of course, with the virtual neutrino. It reduces to the question whether the virtual neutrino can at all be exchanged between the two identical weak vertices. This is a totally symmetric situation which a totally asymmetrical neutrino can not satisfy. It is actually equivalent to the question whether a real neutrino can be captured both by the neutron and the proton [Eq.(3) and Eq(5)], what has been proven negative by the Davis' type of experiments and is reflected in the conservation of the lepton number. In the SM, which has accepted this and the massless Dirac (Weyl) two component neutrino in full, the neutrinoless decay is thus strictly forbidden. It is now only a matter of the extent to which the neutrino still can have properties opposed to those ascertained by present measurements, which can be accommodated into their (comparatively large) experimental incertitude and be responsible for the eventual (small) amplitude of the neutrinoless decay. There are two possibilities for this. The first arises from the still open possibility that the electron neutrino is after all a Majorana particle of small but non-vanishing mass, and the second from the possible admixture to the weak interaction of charged weak currents of opposite handedness, which should be quite small at low energies (if it exists, the 0ν2β decay will, like the proton decay, be a low-energy echo of high-energy symmetries assumed by GUTs). It is obvious that in any case the symmetry of the problem requires the neutrino to be identical to its anti-particle and that, in order to match helicities in both vertices, the neutrino has to have both helicities as well, what can be accomplished either by a massive neutrino, to the extent proportional to its mass $<m_\nu>$, or by the admixture η of the "wrong" current to the interaction. By $<m_\nu>$ here is denoted the so called effective mass of the Majorana neutrino which should be responsible for the effect in the case of neutrino flavor mixing, and which, due to dependence on the phases of mixing coefficients, may turn to be even smaller than any of the component masses. Since for the Majorana neutrino, which is in all quantum numbers identical to its antiparticle, the lepton number can be defined only to distinguish between their helicities (except as a flavor flag), the neutrinoless decay implies the violation of the thus defined lepton number. In spite of great efforts of many people, and achieved sensitivities undreamed of, the process has not been observed to this day (the most promising of all is probably the decay of Ge-76, where the Heidelberg-Moscow experiment in the low-background conditions of the Gran Sasso laboratory has been using the enormous "source-detector-in-one" made out of germanium highly enriched in its isotope 76 just to find out that at the 90%CL the half-life for its neutrinoless decay is longer than $6\times10^{25}$ years. Upper limit for the decay probability and the model values of the nuclear matrix elements define in the η-$<m_\nu>$ plane the region of allowed values whence an upper limit for the Majorana neutrino effective mass of 0.2 eV has been inferred (for future plans under the name of GENIUS, aimed at reaching the sensitivity of 0.01 eV, see Refs. 26 and 37)). Since this pertains only to Majorana neutrino, however, it can not be directly compared to the results deduced from other processes. We have not discussed the relations between different types of the neutrino in any detail, for which we refer the reader to e.g. Ref. 22 or 23, and the discussion of some other relevant details we postpone until we review the Yugoslav activities in this field. With this we close this short review of the neutrinoless decay and continue with the discussion of the basics of the broad field of neutrino oscillations.

a.1.2.2. Neutrino oscillations

Neutrino oscillations were for the first time discussed by Pontecorvo in the distant 1957, shortly after Pais and Piccioni considered similar phenomenon for the neutral kaon. Quantum mechanics of the phenomenon is "somewhat subtle" and we refer the interested reader to specialized treatises for that, e.g. Ref. 24. We first discuss the so called vacuum oscillations of neutrino flavor. If the neutrinos have mass and their flavors mix, then their components of different mass during propagation *in vacuo* disperse and change their relative phases so as to coalesce into another flavor combination at the next weak vertex. If the neutrino was created in a certain vertex as a neutrino of given flavor and of definite mass composition, as determined by the value of θ in Eq.(9) in the case of two flavor mixing, it would, depending also upon the difference of the masses squared of the mass components, $\Delta m^2$, as well as on its energy E, appear at a second vertex L meters away as an admixture of neutrino flavors [28]. For a given set of these parameters the probability to find flavor μ if at creation only flavor e was present is

$$P_{\nu e \to \nu \mu}(E, L, \Delta m^2, \sin^2 2\theta) = \sin^2 2\theta \times \sin^2(1.27 \, \Delta m^2 \, [eV^2] \, L \, [m] / E \, [MeV]) \quad (11)$$

This probability thus oscillates with an "oscillation length" λ, which is the shortest distance between the emission and absorption vertices at which the neutrino initially of one flavor is again its pure own self and which, if measured in kilometers, equals 2.5 E [GeV]/ $\Delta m^2$ [eV$^2$], and does not depend on the extent of the mixing but only on the mass difference between different mass components. Amplitude of the probability, on the other hand, which is $\sin^2 2\theta$, depends on the extent of the mixing only. At its maximum, for θ=45°, at L=λ/2 the initial electron neutrino would interact as the pure muon neutrino. Flavor oscillations obviously violate the conservation of flavor lepton numbers, though the total lepton number is unaltered by oscillations. If oscillations are not observed at a given distance L and for an energy E then the corresponding limiting value of oscillation probability P through Eq.(11) determines in the ($\sin^2 2\theta$, $\Delta m^2$) plane the region of their allowed values (or excludes certain regions). This is how the oscillation experiments are usually analyzed. Different experiments then should yield consistent values for these parameters, or the allowed regions must overlap. In 1985 Mikheyev and Smirnov, following Wolfenstein, developed the idea that the oscillations of neutrino flavors might be resonantly enhanced on their way through matter due to their coherent elastic forward scattering off electrons, which adds to the vacuum propagation phase changes (the MSW effect). The neutrinos of all flavors scatter via neutral current interactions while only electron neutrinos scatter additionally via charged currents. The common neutral current scattering influences only the mass differences, $\Delta m^2$, leaving the mixing angle at the vacuum value $\theta_v$, while the charged current scattering influences the mixing angle only, changing it to the new matter value $\theta_m$. The two mixing angles are related as:

$$\sin 2\theta_m = \sin 2\theta_v / [(\cos 2\theta_v - \lambda / L_o)^2 + \sin^2 2\theta_v]^{1/2} \quad (12)$$

where $L_o$ is the interaction length for the charged current scattering, which is inversely proportional to the electron density of the medium. In general the otherwise small vacuum oscillation amplitudes may thus be greatly amplified if the neutrinos were to pass through the adequate density of matter on their way. At the extreme, for $\lambda / L_o = \cos 2\theta_v$, or at the density of matter called critical, for a given energy, the oscillation amplitude reaches a maximum, $\theta_m=45^o$, meaning that then the initial electron neutrinos may completely change into muon neutrinos. This is likely to happen to solar neutrinos, which on their way from the center of the Sun to its surface see the electron density changing many orders of magnitude, and it is for solar neutrinos that the MSW effect is most seriously considered. It is also possible that the MSW effect to some extent influences neutrinos while they travel through the Earth (though one should not easily dismiss the objections raised in Refs.17, 19). Flavor oscillations occur only between the states of same helicity and both the Dirac and Majorana neutrinos are equally susceptible to flavor oscillations, while only the Majorana neutrino may in principle experience the rather awkward neutrino-antineutrino oscillations, which we shall not comment in any detail.

Two possibilities exist to find out whether the neutrinos indeed oscillate. These are called the "appearance" and the "disappearance" experiments. The first attempt at finding the flavor which initially did not exist while the second would register fewer neutrinos of given flavor than were created. An extremely great variety of these experiments was realized during the period of almost thirty years now, involving neutrinos from reactors, accelerators, our Sun, as well as those created in the Earth's atmosphere by cosmic-ray interactions, this whole activity now bearing the name of "neutrino oscillation industry" (e.g. Ref. 20). Different experiments are sensitive to different ranges of the mass and mixing parameters, as well as to different ranges of the L/E ratio, what is schematically illustrated in Figs. 4a and 4b. Besides, the oscillating low-energy neutrinos (like the reactor or solar electron neutrinos) can not produce the heavy charged leptons and in detectors based on capture reactions may thus only "disappear", while the high energy ones may also lead to observable appearance of new flavors. Detectors based on neutrino-electron scattering, on the other hand, are sensitive to all neutrino flavors, the cross sections being roughly in the ratio of (1:3:6) for the scattering of the (electron neutrino : electron antineutrino : muon/tau neutrino), the flavor asymmetry being due to electron neutrino interacting both by neutral and charged currents and muon/tau neutrinos by neutral currents only. A couple of experiments among all this multitude have recently reported results which may be interpreted as being due to neutrino oscillations and we shall only briefly comment on the most promising of them.

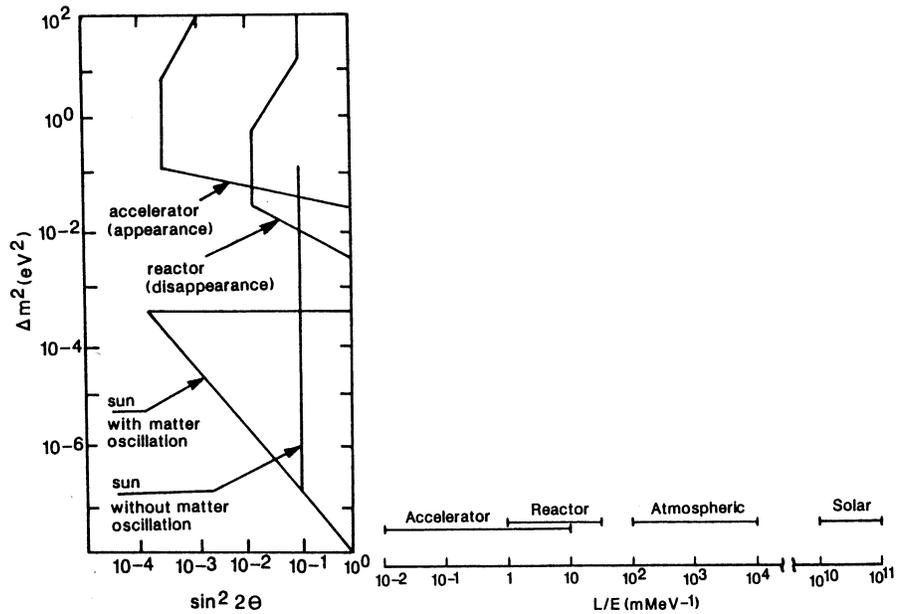

**Fig.4. a.** Values of the mixing and mass parameters accessible to various oscillation experiments (regions to the right of each curve are excluded if at a certain minimum probability oscillations are not observed or are allowed if the oscillations are assumed to be observed), **b.** Values of the parameter L/E accessible to various oscillation experiments [23].

One experiment with accelerator neutrinos, one with atmospheric neutrinos and one with solar neutrinos have up to this day produced results which possibly suggest neutrino flavor oscillations. The accelerator experiment is the Liquid Scintillator Neutrino Detector (LSND) at the LAMPF 800 MeV proton beam stop looking for anti-$\nu_\mu$ oscillations into anti-$\nu_e$ since 1993. On the basis of some 50 to 80 "alien" events altogether (anti-$\nu_e$ + p → n + $e^+$) the oscillation probability of 0.3(1)% has been inferred, suggesting a rather large mass difference for the two lightest neutrinos. This result has, however, not been confirmed in a number of other accelerator experiments and awaits to be either proved or disproved by new measurements.

The mean free path of the high energy hadronic component of primary cosmic rays (mostly protons) in the 10 meters of water equivalent thick Earth's atmosphere amounts to some 80 centimeters of water. With about thirteen interaction lengths the atmosphere is thus a total absorber of the primary spectrum and the source of rich hadronic showers. Atmospheric neutrinos are produced by the decay of cosmic-ray secondaries, kaons, pions and muons, resulting in the muon neutrinos being almost twice as abundant as the electron ones. Most of the detectors built to search for proton decay were having those high-energy neutrinos as a component of background and have been from the early days observing somewhat less muon neutrinos than expected. The most convincing evidence that something is wrong with muon neutrinos came from the biggest of these detectors, actually the biggest of all the detectors ever built, the Super-Kamiokande. The story of Super-K is in itself worth telling. It is a Cherenkov imaging detector containing 50000 tons of ultrapure water viewed by 11200 photomultipliers of half of a meter diameter, located 1000 meters underground, able to distinguish between the high energy electron and muon neutrino events on the basis of distinguishing electrons (which also emit braking radiation and produce less sharp Cherenkov rings) from muons, produced in inelastic scattering off nucleons. The threshold for elastic neutrino scattering (of all flavors) is low, at some 8 MeV, what is determined by background only. Apart from high energy atmospheric neutrinos it thus registers the solar neutrinos as well, and was, due to its directional sensitivity, able to produce the yet only image of our Sun in "neutrino light" (Fig.5). It was wrecked during the writing of this review, on the 12-th of November 2001, in a catastrophic chain-reaction of photomultiplier implosions, when some 7000 of them were destroyed, beyond immediate repair. In 1998, two years after its start, the Super-K team reported the finding that the up and down coming high energy electron neutrinos, as expected for an isotropic situation, produce equal number of electrons, while the up-coming muon neutrinos produce far less muons than the down-coming ones. Since the up-going atmospheric neutrinos pass through the whole Earth their paths to the detector are very much longer than that of the down-going ones and their deficit may be ascribed to oscillations of muon neutrinos into the tau neutrinos, to which the detector is not sensitive (or into some other sterile sort, if it exists). The mixing angle of $\sin^2 2\theta > 0.8$ and mass difference of about $3 \times 10^{-3} eV^2$ fit very well the L/E distributions of muon neutrino events. However, other causes for this deficit have been suggested, among which the very consistent one is the assumed neutrino decay [18]. Sure enough, stability of neutrinos has been questioned in many ways, but only for solar neutrinos on great decay lengths, while for muon neutrinos on decay lengths of couple of kilometers at most. It turns out that the Super-K data are almost equally well reproduced by assuming that the flavor neutrinos are almost pure mass eigenstates, with the $\nu_2$ mass being the greatest, so that the muon neutrino is allowed to decay into the tau neutrino. The ratio of the lifetime to mass of such a neutrino has to be of the order of $10^{-10}$ s/eV. It is interesting that this scenario explains other potential oscillation results as well, the LSND data described above and the deficit of solar neutrinos, while the interpretation in terms of oscillations produces inconsistent set of parameters.

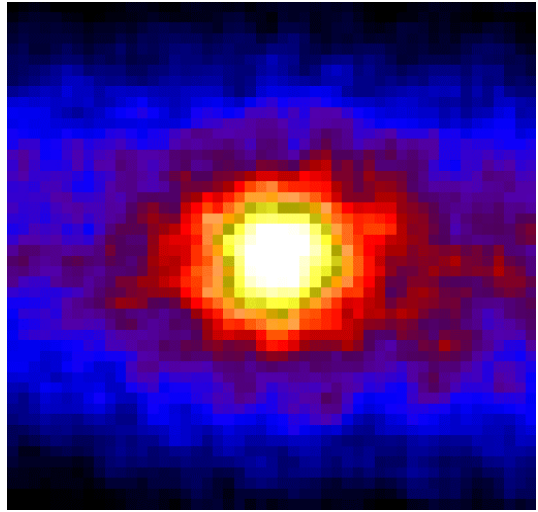

**Fig.5.** Our Sun in neutrino light, as seen by the Super-Kamiokande. The scale of brightness acts as the "neutrino thermometer" of the interior of the Sun.

Finally, the measurement of solar neutrinos produced perhaps the most reliable evidence of neutrino oscillations yet. The problem of the missing solar neutrinos we have already discussed. After the Davis' chlorine experiment, the gallium radiochemical experiments (the GALLEX and SAGE), as well as the Super-K (in different parts of the spectrum, see Fig.3) all found similarly smaller neutrino flux than that predicted by the SSM [30]. This, however, is not necessarily a problem in neutrino physics. There are many aspects of the SSM which are based on insufficiently safe extrapolations and which may perhaps lead to somewhat higher estimates of the neutrino fluxes, though the engineers of the model maintain that the measured fluxes are certainly out of the reach of the model. This is how the problem has been focused on the properties of the neutrino. All the conceivable reasons for the neutrino created in the Sun not to be registered by our detectors require the neutrino to be massive. The possibly decisive measurement of solar neutrinos which might have resolved the problem and at the same time established that the neutrinos posses mass is the result reported in July 2001 by the Sudbury Neutrino Observatory [21]. At the heart of the SNO, which is located at the depth of 6000 m.w.e., there is a 1000 ton of ultrapure heavy water imaging Cherenkov detector, surrounded by 7000 tons of background reducing ordinary water, all this viewed by 9500, 20 cm, photomultipliers. This enables the detector to be sensitive to all neutrino flavors via the elastic neutrino scattering in much the same way as, say, the Super-K, though with significantly smaller efficiency, and additionally, via the charged current reaction $\nu_e + d \rightarrow p + p + e^-$, with much larger cross-section, to electron neutrino only. The SNO relevant data thus consist primarily of this charged current events which may be conveniently compared to the statistically much more significant Super-K data on elastic scattering off electrons. During the 240 days campaign the SNO detector recorded 1170 neutrino events and found the pure electron neutrino flux to be by 3.3 standard deviations (at the 99.96%CL) smaller than the Super-K flux of neutrinos of all flavors, this being the result which allows for interesting interpretations. Firstly, quite independently of the details of the SSM, it says that neutrino flavors other than the electron one contribute to our elastic scattering signals, what testifies that the neutrinos change their flavor on their way here. Quantitatively, it turns out that the solar neutrino flux which reaches the Earth is composed of only about one third of the electron neutrino flavor, the rest two thirds being somehow divided between the other two flavors (all this goes only for the high-energy part of the solar neutrino spectrum in Fig.3, mostly for the boron-8 neutrinos). Secondly, since the absolute electron neutrino event rate is also about one third of what is predicted by the SSM, this at the same time ascertains that the total neutrino flux actually agrees with the SSM. Also, this excludes the possibility of oscillations into any sort of sterile neutrinos. The issue of neutrino oscillations, however, can hardly be considered settled. It remains puzzling why it was not possible to reconcile all the earlier neutrino data by a unique set of oscillation parameters and obviously more systematic research and controlled experiments are needed to clear the matters completely.

a.2. Electromagnetic properties of the neutrino

It remains for us still to briefly discuss the possible electromagnetic properties of the neutrino. Like its other properties this presents a problem of considerable complexity, as is nicely exemplified by the recent review [39]. Being electrically neutral the possible electromagnetic interactions of the neutrino may arise only through higher order diagrams, involving for instance the charged current loops, bearing the name of radiative corrections. A massless neutrino of any kind can not have any electromagnetic moments. When static electromagnetic moments are concerned the CPT invariance, which states that the dipole moments of a particle and its antiparticle must be equal and opposite, forbids even the massive Majorana neutrino to have such moments. Thus, if neutrino were to have a magnetic dipole moment the neutrinoless double beta decay would not be possible, and *vice versa* - the two exclude each other. Non-diagonal, or transition moments, however, may exist for both Majorana and Dirac massive neutrinos, what would lead to their radiative decays between flavors. In the simplest extension of the SM the magnetic moment matrix for the massive Dirac neutrino has elements proportional to m$_{ll'}$, with $l,l'$ =e,μ,τ :

$$\mu_{ll'} [\mu_B] = 3.2 \times 10^{-19} \, m_{ll'} \, [eV] \qquad (13)$$

This is a very small value, far beyond sensitivities of our present-day experiments and leads to no significant effects in neutrino interactions. Certain theories, however, predict that neutrinos could have magnetic dipole moments as large as $10^{-11}$ μ$_B$, what would be almost within the reach of our experiments, which derive the value of this moment from the measurements of cross-sections for elastic neutrino-electron scattering (weak and electromagnetic amplitudes do not interfere and cross sections simply add). If the dipole moment would be this large it would have important effects on neutrino interactions. The most spectacular consequence would be the possibility for the neutrino to flip in a transverse magnetic field from the state of one helicity into the other, sterile helicity state, what could account, say, for the deficit of solar neutrinos. This we shall still comment when further discussing the solar neutrino problem.

In the end, let us sum up this short review of the current status of our knowledge of neutrino properties. The most general conclusion may read that, after almost fifty years of hard work and many thousands of man-years spent, we have not come anywhere near the final answer concerning any one of the neutrino properties. To reach a consistent set of even single figure values which would describe all the neutrino properties is a complex multiparameter problem which would need many more experiments able to control the parameters in a wide range of values. We are perhaps only at the beginning of this deceitfully simple task.

b. Neutrino in branches of knowledge other than neutrino physics

Up to now we tried not to discuss the significance of the neutrino outside the realm of neutrino physics itself, e.g. when we discussed the solar neutrino we did not comment on its significance to the physics of the Sun, or to stellar astrophysics in general, but discussed it only as far as it concerns the properties of the neutrino itself. Now we briefly comment on these other aspects of neutrino physics. Neutrino is to a different degree of interest to cosmology, to astrophysics, and to geophysics, and we will proceed in this order.

Like all the other particles the neutrinos were supposedly primordially created out of vacuum in the Big Bang. Though the details of this most important and complex of models will still vary and evolve together with the evolution of our knowledge some of its general features are perhaps already sufficiently well defined. When the neutrino is concerned these probably amount to the following. During the early phases of the expansion of the Universe its different components are in common thermal equilibrium until for a given component the interaction rate remains much higher than the expansion rate (which is equal to the Hubble constant H, which is in turn proportional to the square root of energy density, $\rho(t)^{1/2}$). When this is no longer the case this particular component decouples from the rest of matter of the Universe and continues to expand and cool down on its own. Due to their weak interaction ability the neutrinos are first to decouple. Different scenarios predict different outcomes, the standard GUT model estimating for the neutrino the decoupling to has happened at about 1 second after the explosion, at the temperature of about 1 MeV. After this instant the total number of those primordial all pervading neutrinos would not change appreciably; they would gradually dilute and, stretching their wave-functions together with the expanding space, cool down to reach in the present epoch, if they were massless, the temperature of some 1.9 K and density of almost $10^9$ per cubic meter, second only to that of the cosmic microwave background photons. Being of such low energy (and if having mass being of even lower energy) those relic neutrinos have desperately small cross-sections and it is not likely that they will ever be directly observed. This is a pity, for it would be good to see this complex debris from the early moments of creation which would, contrary to the featureless decoupled photons, by way of its particle-antiparticle composition tell us something about the origin of the matter-antimatter asymmetry of our world. If, however, they have mass, they can significantly influence the dynamics of the Universe. There is a number of well known arguments which require the existence of mass beyond that connected with luminous matter. This so called dark matter is missing in galaxies to account for their observed rotation velocities, as well as in the Universe as a whole to account for its strongly suggested Euclidean metric. If we assume that whole of this required dark matter is neutrinos than it follows that the mean neutrino mass can not be higher than some 4 to 17 eV, depending on the model applied. If there are more sorts of dark matter, what is quite likely (maybe including the yet unobserved fields, like the recently proposed "quintessence"), this cosmological restriction on neutrino masses gets weaker. If the mean neutrino mass were 30 eV or greater most of the mass of the Universe would be neutrinos! Another large-scale observable sensitive to the properties of neutrinos is the cosmic abundance of helium-4 in the present epoch, which should not be very much different from that originating from the primordial nucleosynthesis in the early Universe [12]. At the moment when the neutrinos decoupled from the rest of matter the neutron to proton ratio stabilized, for their neutrino induced transformations ceased, only to slightly decrease due to neutron decay until when at the temperature of about 100 keV, some 100 seconds after the beginning, the direct synthesis of protons and neutrons into He-4 could have taken place. The expected mass of this primordial He-4 is about 1/4 of all baryonic matter, in accordance with observations. Now, this amount depends, among other things, on the number of neutrino types. This is due to the dependence of the expansion rate of the Universe on the number of neutrino types, the greater the number of neutrino species the greater the expansion rate ($H \propto \rho^{1/2}$), meaning that with the greater number of species the decoupling would have taken place earlier, at higher temperature, when the neutron to proton ratio was higher, what would in turn lead to greater abundance of He-4. Such analyses suggest that the number of neutrino types is between three and four, in surprising agreement with contemporary knowledge. Finally, observation of extragallactic neutrinos and their composition would yield valuable information about the matter-antimatter composition of the Universe at large of which we are otherwise bound to have only indirect information.

When astrophysics is concerned the neutrino plays an important role in the energy balance (or perhaps better say imbalance) in stars in different stages of their evolution. Besides, it serves as a sole uncorrupted carrier of information about the energy generating and isotope creating processes which go on inside the stars during both the stages of quiet and explosive stellar nucleosynthesis. Our Sun is presently in its quiet equilibrium epoch presumably slowly synthesizing helium out of hydrogen in a number of reactions which, to supply the necessary and otherwise non-existent neutrons, must involve weak reactions, which generate neutrinos as well. These low-energy reactions proceed at such a low rate that it has up to now been impossible to study them in the laboratory. The already mentioned SSM is based on extrapolations of our high-energy knowledge and the study of neutrinos emerging from these reactions in the Sun's interior is the only way to check our theories of stellar nucleosynthesis. This is why we would like to measure the entire spectrum of our Fig.3 and why we are so troubled by the disagreement of the SSM and solar neutrino measurements. Our present attitude is that the SSM is right and that the reasons for the disagreement are to be sought elsewhere. We have already mentioned a number of possible causes devised to save the SSM, which would be due to the properties of the neutrino, and have hopefully demonstrated how difficult it is to single out the most likely one. Now, for the sake of completeness and to add to the confusion, we mention yet another one in this category. It has to do with the observation by the chlorine experiment of the possible 11-year periodicity of the measured neutrino flux, which is found anti-correlated with the well known variation of solar activity of the same period. Though the statistical significance of this correlation is a matter of controversy it can not be easily dismissed, for if true it would bear heavily on the SSM. Among other possibilities, if the neutrino were to have a rather large magnetic moment, this offers a solution to the solar neutrino problem, for its spin flip during the large amplitude values of solar magnetic fields would turn them sterile to our capture detectors, and more difficult to detect by our scattering detectors. Another awkward possibility we save for our story of the thallium solar neutrino detector. The neutrinos which reach us from the Sun thus dissipate a significant part of the energy generated in the exoenergetic nuclear processes, which therefore does not contribute to the equilibrium of pressures, what somewhat accelerates all the processes which go on. In very much heavier stars the nuclear fuels are used up at a very much higher rate and after the maximum binding energy per nucleon is reached, under the enormous pressures the endoenergetic processes which form very heavy nuclei start to wind up. For these nuclei with higher neutron to proton ratio to form more fresh neutrons are needed and these are provided by forced electron captures which generate the neutrinos as well. All this consumes energy at a very high rate and leads to rapid contraction which is greatly accelerated by the escaping neutrinos. It is for this effect of the neutrinos that Gamow has coined the term "Urca process" (after the name of the casino in Rio de Janeiro, which drains the money out of one's pocket as fast as the neutrinos drain the energy out of the collapsing star). This results in the implosion of a star which is then followed by what we call the explosion of the supernova. This (greatly simplified) scenario leads to a powerful neutrino burst at a certain stage of the explosion. Some of our big proton decay detectors registered a couple of neutrinos from the nearby supernova SN1987A in February 1987. The delay of this neutrino pulse with respect to the optical registration of the explosion, and its spread in time, yields the upper limit for the electron neutrino mass of the order of 10 to 30 eV. Neutrino astrophysics of the future will no doubt aim at exploiting similar events to a much greater advantage.

Finally, we comment on possible interest of geophysics and geology in neutrinos. When the knowledge of our Earth's interior is in question the notorious fact is that it is completely inaccessible to direct examination and that we have, much like with other heavenly bodies, to rely on indirect evidence only. An important thing, however, we know for sure; its volcanic activity brings about the fact that its material is still hot and molten under the thin crust. Besides, geological differentiation of the crust witnesses that it has been liquid once so that the elements might migrate and form great mineralizations. Exact dating methods agree that the mineralizations occurred long after the Earth would have congealed if there were not for some low power but long lasting source of energy, other than the quickly (in some tens of millions of years) consumed gravitational contraction. From the early days of radioactivity it has been clear that natural radioactivity might have kept the Earth's crust liquid long enough and that it probably keeps the Earth's interior liquid even today. Direct evidence, however, does not exist; every addition is welcome and neutrinos may perhaps help in the following. Glashow, Krauss and Schramm have estimated that natural radioactivity in the present epoch should yield an antineutrino flux on the surface of the Earth of about $10^7$ neutrinos per $cm^2$ per second and that therefore constitutes an non-negligible part of the neutrino bath we live in. If it could be measured it would help to reliably know the amount of radiogenic heat and to understand our home planet better. Atmospheric neutrinos, on the other hand, may hopefully be used for the "neutrinography" of the Earth, or perhaps even for the "computerized neutrino tomography" of the Earth which might yield the detailed map of its interior. With this we conclude our short review of the role of neutrino in cosmo-, astro- and geo- sciences, which mostly belong to the emerging field of astroparticle physics, and move on to a brief summary of the status of neutrino physics today.

## A.2. The present status of neutrino physics

Having explored the most important intricacies of the physics of the neutrino we may rephrase our earlier conclusion about its present status - that practically all the dilemmas we have ever had about the neutrino survive to this day! Some progress is, however, perhaps in sight; the mentioned indications of flavor oscillations [21], [34], and then the recent suggestion that neutrinos after all may interact somewhat differently than expected, the Weinberg angle from neutrino interactions being slightly but significantly different from that obtained from other interactions [40] (but also see [41]). The many experiments which are either planned or are soon to start mark only the beginning of the new generation of great experiments aimed at further narrowing the values of neutrino parameters (the long-base oscillation experiments: KEK to Super-K (which now has to be modified), MINOS experiment (FNAL to Soudan, at the base of 730 kilometers and neutrino energy of 10 GeV) and CERN to Gran Sasso, at the base of similar length; the short-base MiniBooNE experiment at FNAL to test the LSND result; the GENIUS Ge-76 neutrinoless decay experiment of incerased sensitivity, the extension of SNO measurements to neutral current interactions of solar neutrinos, or the BOREXINO, the 300 tone liquid scintillator in Gran Sasso to measure the Be-7 solar neutrino flux, to mention just the few).

## A.3. The future of neutrino physics

It is perhaps still true that, in spite of all the necessary heavy planning of our experiments of today, physics truly progresses only through surprises - and since surprises can not be predicted, one is bound to search for them, guided more by intuition and experience than by logically closed arguments based on existing knowledge. Because experiment is to find and theory is to explain, but again theory is to suspect and experiment is to confirm, surprises occur equally in theory as in the experiment - everybody has been utterly surprised by the discovery of relativity, Einstein included, or, on the other hand, by the discovery of the muon, Anderson included. Even the smallest and most insignificant of discoveries must contain an element of surprise if it is to fall into this praised category which enriches our knowledge most. Now, discoveries in theory and experiment must, with fluctuations, necessarily alternate and if experiment is to keep pace with theory, or if evolution of knowledge is to continue, purely experimental search for surprises must go on, in much the same way as it always freely goes on within the sphere of the mind. We have to believe that time of miracles is not in the past and that the more we work the more miracles we shall encounter. This small exercise in logic was meant only to excuse us from obligation to predict the future of neutrino physics, what is more often than not an unrewarding business.

Some general aspects of neutrino physics in the future are, however, predictable. Though many varieties of Grand Unified Theories offer different possibilities for the neutrino (e.g. the possibility to have the light and heavy neutrinos via the see-saw mechanism, etc.) it is not likely that we shall have any firm theoretical guidance in our future investigation of neutrino properties and it is pretty certain that actual ambiguities are to be solved only experimentally. We have already issued the most important caveat concerning the future of neutrino physics, namely that, if the Standard Model is right, it will take an infinite time to prove it so! In that case painstaking systematic search through the multidimensional space of neutrino parameters in the controlled accelerator experiments, as well as the perpetual increase of sensitivity of non-accelerator low-energy neutrino measurements, will lead to asymptotic closing on the null values of neutrino parameters. If, on the other hand, the neutrino parameters have finite values we shall sooner or later find them and set the firm basis for a new theory. New ideas to build the neutrino factories are being developed; very high intensity neutrino beams may be produced by muons created from an intense pion source at low energies, their phase space then compressed to produce a bright beam which is then accelerated to the desired energy and injected into a storage ring with long straight sections pointing in the desired direction, or, focused quality beams of electron anti-neutrinos may be produced by accelerating and storing the beta decaying nuclei [46]. All this, however, remains within the limits of the existing accelerator technology.

As suggested by perusing the bulky volumes of the Proceedings of the last International Cosmic-Ray Conference [45] (some 4500 pages altogether!) another probable development is the rise of cosmic-ray based Extra-High-Energy neutrino physics. One can not help the impression that the global net of EHE experiments is already gradually being built, aimed at capturing exclusively the extremely rare events produced by EHE cosmic-rays not only in the Earth's atmosphere, but in the biggest targets available, Earth itself, or even the Moon, at energies orders of magnitude beyond those available at the accelerators; and that at those nearing the GUT energies cosmic-rays are to take over the lead back from the accelerators. That the measurements of the neutrino component of such events would be essential to their understanding, as well as to extracting information about new physics of the neutrino itself, is beyond doubt. The so-called natural Cherenkov neutrino detectors (Baikal, ANTARES, AMANDA, etc.) are good first steps in this direction. Connections between the accelerator and cosmic-ray EHE physics are already being developed in the overlaping energy regions, where the accelerator data are used for the simulation of cosmic-ray events, in order to check the quality of simulations extrapolated into the regions of future EHE cosmic-ray physics.

To avoid an anti-climax this is where we end our review of the past, present and future of neutrino physics and move on to discuss its few relations with Yugoslav physics.

**B. THE YUGOSLAV CONNECTION**

Yugoslav physics has always been a small, if not to say poor-man's physics (what is not in itself derogatory, and is due to causes beyond physics) meaning that it has never had neither big accelerators nor big detectors, and that even when practiced in comparatively big institutes it nourished no great endeavors. It is almost a miracle then that it has had any connections whatsoever with the field as demanding and costly as neutrino physics. That there is anything at all worth mentioning under the title above is due to a number of haphazard circumstances and due to a number of people able to recognize a chance when they stumble upon one, and then willing to attempt the impossible. This story of Yugoslav connections with neutrino physics may be of interest to others in similar position as well as to the physics youth and future of physics in Yugoslavia itself and this is why we set to prepare this short review. There are two fields of neutrino physics to which Yugoslav physics modestly contributed - to the studies of the solar neutrino and to the study of the neutrinoless double beta decay.

**B.1. The Thallium solar neutrino experiment**
*Thallium as the lowest energy solar neutrino detector and monitor of the solar neutrino flux in the past - the closest encounter yet of nu and yu physics*

It was in the early days of the solar neutrino problem, when all the possibilities to independently check this irritating result were systematically explored, that thallium has been found suitable as a geochemical solar neutrino detector of quite distinctive properties. This has been practically an one-man job, that of Melvin S. Freedman (1916-1997), then, as during the all of his more than fifty fruitful active years as a nuclear chemist (sic!), at Argonne. The proposal appeared for the first time in 1976 [42], but the first comprehensive study he presented at an informal conference on solar neutrino research held in Brookhaven, already in 1978 [43]. There he touched upon practically each of the many problems from vitrually every science imaginable pertaining to this difficult measurement and gave hints as to their solutions. The difficulties are perhaps best apprehended by noting that the measurement is, 25 years later, not yet done. To make this long story short we have to dwell mainly on the details of interest to our purpose and leave out the mention of many important contributions made to the project during all these years. We apologize for that.

We first make the case for the thallium experiment. The thallium detector is based on the capture of the neutrino, of energy higher than about 50 keV, by its isotope Tl-205 (70%), transforming it into lead-205, which then by electron-capture decays back to Tl-205 with the half-life of some 15 million years. The content of Pb-205 in the thallium mineral, presumably well isolated from the rest of the world, will thus saturate after some tens of millions of years at the level proportional to the capture probability averaged over the neutrino spectrum above the (low) reaction threshold (see Fig.3) and the neutrino flux averaged over the exposure time. Measuring the lead-205 content and knowing the capture probability would yield the average neutrino flux. Now, except for the most general of arguments, which is due to what may justly be called the golden rule of experimental physics No.1, which states that no thing is known unless the results of at least two measurements based on widely different principles (with different systematic errors) agree within the cited uncertainties - which is perhaps in neutrino physics more truly so than anywhere else and justifies the need for as many different experiments as possible, there is an argument supporting the thallium experiment in particular. At its extreme it is nicely illustrated by the witty remark of William Fowler's, worthy of the great connoisseur, made in the days when the first results of the Davis' experiment suggested that virtually no neutrinos from the Sun are being registered by the chlorine detector. Fowler offered an explanation (seemingly jocular) stating that the Sun is not only maybe dying (in the thermonuclear sense) but that it has possibly been dead for millions of years now (to appreciate this one has to be aware that for the electromagnetic radiation generated in the center of the Sun it takes some ten million years to reach its surface, degraded to the 6000 K equilibrium spectrum, and that if the Sun was to die today this would not affect its surface radiation for another couple million years, while the neutrino emission would cease immediately). That would not indicate such a bad physics on our behalf as it seems; it would mean that we have misjudged the lifetime of the Sun by about the same factor by which we have misjudged the neutrino flux, though, of course, it would be a nasty coincidence for the Sun to start to change its output just when we managed to measure it (within, say, one part in thousand of its lifetime) (to illustrate the state of minds at that time, however, the present author recalls himself suggesting in despair that we perhaps deal here with an anti-sun!)). By measuring the average solar neutrino flux in the past the thallium experiment will thus assure us that everything is in order with our Sun, and that we are still left sufficient time to find out how it really works (another proposed geochemical detector, the Mo-98, which also yields a long-lived isotope, has a very high threshold and is not sensitive to basic pp neutrinos).

We now discuss the three main problems of the thallium experiment in the order in which they had to be answered during the feasibility study, which is however perhaps not yet completed (!). At the same time we shall try to tell the story of the Project, which is in more detail told elsewhere by M.K.Pavićević, its main actor on the Yugoslav side [50].

1. The first is the problem of the source material. Thallium is a dispersed element and its mineral of well-known age, sufficiently well isolated from all the many possible sources of contamination with lead-205, and in sufficient quantity, is not known to exist. The ancient arsenic mine in the mountainous uninhabited region in Yugoslav part of Macedonia, known as the Allchar mine, near the border with Greece, with its comparatively large supply of the thallium mineral lorandite, comes closest to those requirements, but whether close enough is not clear yet. Being located in the former Yugoslavia, this is how Yugoslav physics ultimately got involved with the thallium project.

There is, however, another episode involving the same mineral and the mine, which preceded this one, and which significantly influenced the course of events of our present interest. At about the same time when the solar neutrino problem emerged a number of groups around the world were pursuing the search for the long-lived isotopes of superheavy elements in nature, which would by assumption belong to the, at that time yet beyond the reach of our accelerators, island of stability around the theoretically predicted closed shell at $Z=114$. The $Z=113$ element, being chemically homologous to thallium (the eka-thallium), is expected to follow thallium in geochemical processes and its long-lived isotopes are expected to be found, if at all, within, or close to its minerals. This is how, in the early seventies, the well known Dubna group of G.Flerov for the first time approached the Yugoslav authorities, asking to analyse the minerals from Allchar for presence of eka-thallium, in collaboration with Yugoslav scientists. S.Ribnikar from the Faculty of Physical-chemistry in Belgrade was allotted to lead the collaboration from the Yugoslav side. Submitting the proposal, however, the scientists committed a serious mistake; to support their cause better, among other arguments they suggested that the long-lived isotopes of superheavy elements might have extremely small critical masses for the fission chain reaction. This turned out to be fatal not only to this project but has had serious aftereffects on the thallium solar neutrino project as well. The committee of the project declared it to be of special interest to the country and put the project, mine included, under the control of the army. This eventually stopped the project [1].

Thus, when Freedman and his collaborators approached Yugoslav authorities, this has been for the second time within a short interval, and that aroused suspicions about the foul play on behalf of the great powers (what was perhaps a rather natural reaction of the laymen) and many obstacles were implicitly raised on the way to proper uses of the mine. This, however, has not been the reason for the suspension of the project in the USA. As we shall see in more detail later on, it was John Bahcall who expressed one serious objection to Freedman's proposal, stating that the estimates of neutrino capture probability are not sufficiently reliable [15], what was probably the main reason why the ANL proposal to the NSF, submitted in 1977, has been turned down. The idea then laid abandoned for some five years, to be revived in Germany in 1983. The initiators were H.Morinaga, then at TU Munchen, W.Henning, then at Argonne, and the group from GSI Darmstadt, who suggested the emerging Accelerator Mass Spectrometry (AMS) as the method for detection of minute quantities of lead-205. By the strange coincidence M.K.Pavićević from the Faculty of Mining and Geology in Belgrade, who was then visiting at the Max

---

[1] The only endeavor in this direction, which produced numerical result, was conducted by R.Antanasijević at the Institute of Physics in Belgrade. His group built the large NE343 gadolinium enriched liquid scintillator spherical detector, one meter in diameter and 600 liter volume, viewed by twelve 15 cm PMTs, what under the circumstances was a major undertaking, which was to look for spontaneous fission neutrons of high multiplicity, what was supposed to be one of the decay modes of the SHE which would live long enough to be still found in nature. To reduce the high background the digital storage oscilloscope which registered the scintillator pulses was triggered by the silicon detector of fission fragments which were eventually emitted from the ultrathin evaporated sample of the mineral, positioned at the center of the big sphere. The quantity of the mineral was thus small but in return the signature was so stringent that the measurement was virtually backgroundless. This measurement lasted for one whole year. In another measurement the solid track detectors were used, exposed to eventual fission fragments from thin specimens for the period of 10 years. None of the experiments registered a single event which could be interpreted as being due to spontaneous fission of the SHE, what yielded the upper limit for an isotope of eka-thallium with the partial lifetime for spontaneous fission longer than $10^9$ years of $7 \times 10^{-11}$ grams per gram of the mineral. This is also worth mentioning since, unexpectedly, this experiment turned out to be the first to measure the U content in lorandite as well as in its co-genetic minerals, the realgar and the orpiment, the measurement of interest to the thallium solar experiment as well! Due to a number of circumstances, however, the results of this experiment, of an unusual scale for Yugoslav physics, were never published in full [47].

Planck Institute in Heidelberg, has been approached by the colleague from Soviet Union who offered a sample of lorandite from Allchar, which she possessed from the time of the first eka-thallium search, to be inspected by PIXE for the presence of eka-thallium. Pavićević and Morinaga eventually met, exchanged their knowledge about lorandite, found each other believers in the Freedman's idea, and formed an alliance which lasted for some six years during which the thallium project significantly advanced. Mostly due to the inexhaustible enthusiasm of M.Pavićević the Project in 1985 got funded for the first time, and that had to be by the Research Council of Yugoslavia! The funding continued by different Yugoslav Agencies even through the most troubled years of Yugoslav recent past, all the way to the year 2000 (in the meantime the joint German-Yugoslav proposal was rejected by the EC, Brussels). During that period two great international conferences dedicated to the Project (which was in 1986 abbreviated to LOREX, for LORandite EXperiment) were organized by Pavićević, [48] and [49], as well as five international Workshops, the last of which in 1998. Thanks to this uninterrupted effort the Project is today close to its realization, while other potential long-lived geochemical detector projects during this time perished (Mo-98, Br-71).

We now sum up the current status of the problems related to the source material. The age of the mineral from Allchar has been by exact dating methods determined to some 4.2 million years, what leaves the lead-205 concentration rather below saturation. This cannot be helped. The available quantity of the mineral is not yet reliably known. Extensive (and expensive) works in the mine would be needed to reach the deeply located ore-bodies which, according to local miners, contain large quantities of lorandite. If the sensitivity of the detection methods would be increased, even the already available quantity of the mineral might be sufficient. The most serious problem is that of background. There are many ways other than solar neutrino capture which may lead to the presence of $^{205}$Pb in the mineral. Already Freedman discussed majority of the background forming reactions and singled out the $^{205}$Tl(p,n)$^{205}$Pb reaction, the protons being due to cosmic-ray muon interactions in the rock, as the most dangerous one. Only if the mineral has been deep enough for most of the time would this background be acceptably low. Since due to the ubiquitous erosion it must have been deeper in the past than it is today, in the final score this depends upon the so called erosion rate, usually expressed in meters of overburden rock eroded per million years. Different estimates yielded different erosion rates, while indirect muon in-beam measurements performed at CERN by the Munchen group found the contribution to background somewhat higher than expected from extrapolations [51]. Yugoslav physicists proposed the method to measure this component of background directly [60]. The idea is a simple one; if one would know the contents of lead-205 as a function of the present-day depth of the mineral one would, assuming a definite erosion model (the linear one being justified), be able to derive both the muon depth-dependent contribution and the constant bias due to the neutrino capture and other depth-independent background contributions, hopefully with acceptable errors (these other contributions, mostly due to neutron capture on $^{204}$Pb, may be determined independently, by measuring the lead-205 content in the coexisting minerals, the realgar and the orpiment). Thus, if the mineral would be available from a couple of different depths, this could solve all the otherwise serious background problems. If this is realistic remains to be seen. Another Yugoslav contribution also dealt with possible background to geochemical experiments in general [52]. It is evident that if one deals with a low-probability process induced by high-flux of incoming particles (as is the case with the neutrino induced processes) one would get a comparable yield by a background process induced by a negligible flux of incoming particles if the cross-sections would be correspondingly higher. Now, for the solar neutrino induced processes it turns out that if one would have a flux of background inducing particles of only say one particle per second per cm$^2$ then the process with the cross section of the order of some $10^{-35}$ cm$^2$ would already start contributing significantly. No attention is usually paid to such low-probability higher order processes and the authors pointed out that this is why the fluxes of all the environmental radiations should be measured *in situ*, for even the environmental gamma rays (of equilibrium continuous scattering spectrum, stretching to 2.6 MeV) may lead to nuclear transmutations via the succession of processes which could procede via the excitation (both resonant and non-resonant [32], [33] , [36]) of the higher short-lived state in the parent isotope and its succesive electromagnetic decay to the lower longer-lived excited state, which would in turn eventually have some beta-decay branching to one, or both, of its neighbor isobars. Summing through all the possible channels which, depending on the (often unknown) details of the high-energy parts of the decay schemes of all the nuclei involved, can be many, may in favorable (or better say unfavorable) cases lead to the effective cross-section close to dangerous values. This possibility was later named LeGINT (standing for Low-energy Gamma-ray Induced Nuclear Transmutations). It has been demonstrated that in the case of gold the (surprisingly high) cross-section for the LeGINT, as induced by the Co-60 gamma rays, is of the order of $10^{-31}$ cm$^2$ keV [53], while in the case of tin, irradiated by the 4 MeV electron linac bremsstrahlung, it is of the order of 10 nbarns. Yugoslav physicists also organized a small expedition to the mine, carrying the gamma-ray spectroscopic equipment on donkeys' backs, to measure the gamma-ray flux in the accessible part of the mine, as well as the cosmic-ray muon intensity there in order to determine the effective equivalent depth [35] of that particular ore-body. The gamma-ray flux was found to be low and the effective equivalent depth of that shallowest of all the ore-bodies was determined to some 40 m.w.e., but those results were, however, never published. This is about where the direct contributions of Yugoslav physics to the Project end. Many other contributions, however, including the future measurements of $^{10}$Be and $^{26}$Al in samples of quartz from the same locality to be performed by AMS in Vienna (W.Kutschera and M.Pavićević, with very encouraging preliminary results [67]) or at TU Munchen (E.Nolte), in order to determine the erosion rates accurately, should help solving the background problem.

2. The second great problem with the thallium experiment, though historically it was the first one, is that the neutrino capture probability is not sufficiently well known. At its worst the capture rate might turn out too low and the resulting lead-205 concentration below all the possible detection limits, or comparable to, or even below, the background concentration. If it would turn out high, on the other hand, but again uncertain, we would not be able to know the neutrino flux. In this case, however, the measurement would still yield the useful result and we could wait with the interpretation until the cross-sections are known better.

To discuss this interesting problem we have to have a look at the relevant part of the A=205 decay scheme which is, in somewhat unusual form, together with Freedman [55], presented in Fig.6. The neutrino capture by $^{205}$Tl to the ground state of $^{205}$Pb, which is the inverse of the $5/2^- \to 1/2^+$ electron capture with *log ft* = 10.3, and has the corresponding neutrino capture cross section of the order of $10^{-50}$ cm$^2$, is seen to be much less probable than the capture to the first 2.3 keV excited state, which is the $1/2^+ \to 1/2^-$ transition. Now, Freedman found a number of equivalent beta transitions in neighboring odd nuclei, the $1/2^-$ ground state of $^{205}$Hg decay to the same $1/2^+$ ground state of $^{205}$Tl in the first place, with well known *log ft* values in the interval of 5.1 to 5.3, and, by analogy, estimated the transition we are interested in to have *log ft* equal to some 5.3. The corresponding capture cross-section is then expected to be of the quite satisfactory order of $10^{-45}$ cm$^2$, five orders of magnitude more probable than the ground state capture, as suggested by their respective *ft* values. This is where J.Bahcall came in with the remark that all of these analogous transitions are of the prevailing $p_{1/2}$ neutron to $s_{1/2}$ proton character while the states involved in the transition in question are of different single-particle configurations. Inspecting the single-particle states in the vicinity of the Z=82 and N=126 closures shows that the principal proton configuration of lead is $s_{1/2}^2$ and that of thallium is $s_{1/2}^1$, while the neutron configuration of the 2.3 keV state in $^{205}$Pb is $f_{5/2}^4 p_{1/2}^1$ (as compared to the $f_{5/2}^5$ configuration of the ground state) and the principal neutron configuration of the $^{205}$Tl ground state is $f_{5/2}^6$. The transition in question can thus go only as a two-particle transition, what is only second order via weak interaction. Way out is an admixture of the $f_{5/2}^4 p_{1/2}^2$ neutron configuration to the ground state of thallium, when a single-particle transition becomes possible. Freedman used the data from a number of particle transfer reactions involving nearby nuclei to find out that the amplitude of this component is about 0.25, what reduced his initial estimate by a factor of 4, leading to an estimate of the overall neutrino capture rate into the 2.3 keV state of about still very satisfactory 100 SNU, or to the total of some 40 atoms of lead-205 per gram of lorandite. Bahcall, however, had further objections to using the data from particle transfer reactions for this purpose, stating that, contrary to the weak interaction, the strong interaction involved in these reactions may change the relative phases of the participating wave functions to the extent which is not known. In another independent attempt K.Ogawa and K.Arita [57] performed rather detailed shell-model calculation of this transition probability, their value, due to cancellation of some matrix elements, being somewhat lower than that of Freedman. On the other hand, they demonstrated that the neutrino capture to higher states of lead-205 cannot be neglected and that their contribution must be carefully evaluated.

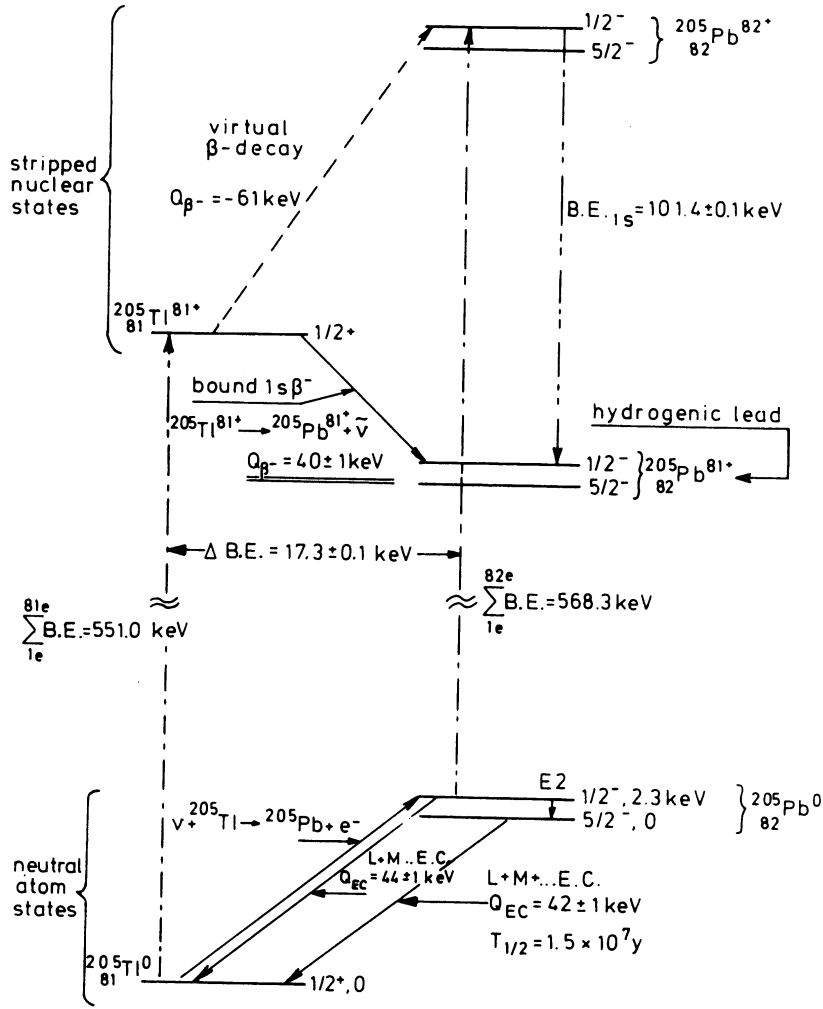

**Fig.6.** The relevant part of the decay scheme of A=205 nuclei [55]. See text for details

The previous paragraph was primarily intended to emphasize the importance of direct experimental determination of the relevant transition probability. That this is at all possible has been for the first time put forward by P.Kienle [58] and elaborated by Freedman [55] more than 15 years ago. The idea involves one of the most exotic nuclear processes, the realization of which borders with the impossible, and is worth telling in some detail. The process in question is the so called bound beta decay of the bare nucleus, stripped of all of its electrons. The bound beta decay differs from an ordinary beta decay only in the final state of the beta electron, which here ends somewhere in the discrete spectrum of the atomic bound states and not in the continuum. In the neutral or weakly ionized atoms the amplitudes of wave functions of unoccupied electron states at the nucleus are very small and the process is unobservable in competition with the common beta decay. In highly and especially in totally ionized atoms the bound beta decay becomes a significant decay mode which can drastically change the lifetime of the nucleus. It has, however, due to obvious experimental difficulties, been for the first time directly observed only last year [59]. Now, within the whole atomic zoo there are only two atoms, thallium-205 being one of them, which are otherwise stable but which, only when fully ionized, decay by the bound beta decay, this then being their only decay mode. This we have deliberately called "the decay of the atom", for it is the best example to demonstrate the often forgotten fact that it is atomic masses, electron binding energies included, which govern the nuclear processes. By another convenient coincidence it was nobody else but Melvin Freedman who was the first to prove in the remote 1952 [66] that the difference between the total electron binding energies of the parent and daughter atom is carried away as a certain excitation energy by the recoiling daughter atom, and not by the emitted radiations. It is thus no wonder that Freedman was ready to deal with the thallium-205 bound beta decay when it appeared of significance to the Project.

A look at our Fig.6 makes some important relations obvious. First, one sees that the $\beta^-$ decay of the bare $^{205}$Tl nucleus (the 81+ ion) is possible only to the hidrogen-like 1s state of $^{205}$Pb (again an 81+ ion) and that it is most probable to the 2.3 keV state of its nucleus. One further sees that this decay is just the time reversed electron capture decay of the 2.3 keV state of neutral $^{205}$Pb leading to the ground state of neutral $^{205}$Tl. This, on the other hand, is the inverse of the neutrino capture we are interested in. Measuring the lifetime of the bare $^{205}$Tl decay would thus yield the matrix element needed to know the neutrino capture probability to the 2.3 keV state. Using three different methods Freedman arrived at an estimate for the decay constant of the fully ionized thallium-205 of $4\times10^{-6}$ h$^{-1}$, what has by P.Kienle been judged measurable by the extraordinary technique which we shall now briefly describe.

At the GSI Darmstadt there has been built the by now well known heavy-ion accelerator complex, the last stage of which is the Experimental Storage Ring (ESR) capable of circulating without appreciable loss the high-energy (but only mildly relativistic) beams of multiply ionized heavy-ions for many hours. If the ions decay and if their decay products are accumulated in the beam for sufficient time, and then counted, the decay constant of the circulating ions can be inferred. This has been done in 1997 with rhenium-187 and similar measurement with thallium-205 is under preparation; the detector setup to count the produced lead-205 ions has been built and tested [38] and the experiment is approved for the year 2002. We may thus expect that the rate of the main neutrino capture branch will soon be known with high accuracy; for the capture rates into higher states we would still have to rely on theoretical estimates.

3. The last major problem of the Project is that the technique to reliably count the anticipated absolute number of atoms of lead-205 is not yet satisfactorily developed.

The relative mass difference of lead-205 and thallium-205 is only 3 in $10^7$, what is well beyond resolution power of standard mass spectrometry, and already Freedman was aware of the need for extremely strict chemical separation of lead from thallium, what proved itself to be a very difficult problem. All other methods which were considered, like the different varieties of laser resonance techniques, or standard accelerator mass spectrometry, more or less suffer from the same problem. Independent of this isobar resolution problem is the problem of absolute sensitivity, or overall efficiency, of the lead-205 detection method, owing to the smallness of which the quantity of the mineral to be processed becomes prohibitively large. One single technique, named the Schottky mass spectrometry, which is uniquely practiced at the ESR, perhaps offers the final solution to both problems. This is the same technique which was used in the mentioned first direct detection of the bound beta decay [59]. In brief it works in the following way. The circulating ions, their charge state controlled by the interactions with the gas target which may be switched on and off, are electron cooled and detected via the time-resolved Schottky spectroscopy. The beam noise induced in the pick-up electrodes is frequency analyzed, the intensity of the line at the revolution frequency of each stored ion species being proportional to the number of ions in the beam. Mass resolution is such that at A≈50 the ions in their nuclear ground state are well separated from those in their nuclear isomer state, if this state is higher than some 100 keV (decay of the intensity of the line belonging to the unstable species measures its lifetime).

Since the storage times may be very long the absolute sensitivity can also be very high, so that even the already available quantity of the mineral might appear sufficient, at the same time the requirements on chemistry also being strongly relaxed. Pilot experiments, which are to take place this year, are being prepared and it will soon be known if this scheme will meet the expectations.

We thus see that the thallium solar neutrino experiment is finally, 25 years after it has been conceived, probably nearing realization. It is a great pity that Mel Freedman will not be around to rejoice, for this is the experiment which should justly bear his name. As it turned out, the realization of the experiment will in the final score probably be another well deserved success of the fully revived German physics. To the Yugoslav side, however, there goes some credit as well. If it was not for the resourceful and persistent Mića Pavićević the LOREX project would probably never come to life, in the first place. If he was not backed up by the Yugoslav scientists the Yugoslav authorities would never have supported the Project for such a long time - the long time that was, as it is now clear, necessary for all the difficult problems posed by the experiment to be solved. If it was not for Mića Pavićević to coordinate, and Yugoslav authorities to support the organization, there would be no international conferences on the subject and there would be no invaluable proceedings to guide further coordinated actions. Now, waiting for the final results, this is where we leave the story of the Thallium solar neutrino experiment and its connections with Yugoslav science and move on to briefly review another connection of Yugoslav physics with the physics of the neutrino.

### B.2. The neutrinoless double beta decay
*Another modest contribution of Yugoslav physics to the physics of the neutrino*

In the mid seventies of the last century, after some ten years of intensive uses of germanium detectors in the detailed nuclear spectroscopy studies of low-lying states of the daughters of long-lived isotopes, largely due to the inability of theory to follow with the same precision, this whole field lost its initial attraction. Having no access to short lived isotopes, it was then that L.Marinkov, who was moving from the Vinca Institute to the University of Novi Sad, initiated there the development of low-background gamma-ray spectroscopy with the aim to study rare nuclear processes in general. Out of 22 tons of the pre-WWII iron he built there the big low-background chamber, of about one $m^3$ inner volume and 25 cm thick walls. In the decades to follow I.Bikit took over and within this passive shielding managed to assemble the germanium spectrometer actively shielded by a big NaI annulus. The low-background performance of the set-up turned out satisfactory [31] and, besides the daily routine with environmental samples which occupied most of the measuring time, the study of double beta decays gradually emerged as a natural choice. Most of these long-lasting measurements which require unperturbed operation and stable conditions were performed during the most hard of times Yugoslavia has been through. Everybody got used to cope with the daily elementary problems with power failures, the instability of the mains voltage, shortage of the liquid nitrogen, lack of small spare parts, and the like, not to mention the embargo on scientific equipment in general. The results which we are about to comment are thus perhaps even more to cherish than it may appear at first sight.

Minding the great efforts and investments into the study of double beta decay worldwide, one had to be careful not to get involved in a highly competitive and demanding field with inadequate means and arrive at only inferior results. Upon the inspection of all the double beta decay cases, and after carefully evaluated the performance and potentials of their set-up, the Novi Sad group of I.Bikit singled out a number of double proton decays as the promising candidates to study. The results on them were scarce and, having a good fast-slow coincidence system of shielded high efficiency detectors well suited to the detection of annihilation radiation, it seemed that it would be possible to improve upon the existing limits on lifetimes of these isotopes. Their experience with low-background measurements, including the discovery of the unknown weak gamma-ray in the decay of Cs-137, an addition to the simplest and presumably best known of all the decay schemes [62], as well as the above mentioned LeGINT measurements and different resonant and non-resonant photoactivation studies, was helpful and encouraging in this respect.

At the root of all the searches for rare processes is the simple expression which, if the decay has not been observed by attempting to detect a certain radiation supposedly emitted in the decay, gives the lower limit of the half-life as:

$$T_{1/2} > const(\text{C.L.})\, \varepsilon\, p\, T_m\, N / B^{1/2} \qquad (14)$$

where $\varepsilon$ is the overall detection efficiency, $p$ is the absolute intensity of the radiation sought, $N$ is the number of atoms susceptible to the suspected decay, and $B$ is the total number of indistinguishable background events accumulated during the measurement time $T_m$. The constant is (non-linearly) inversely proportional to the confidence level at which it has been decided that the intensity sought is smaller than the fluctuation of the background. The trick of the trade is then to minimize background, both by removing all of its possible sources and by defining the event of interest more stringently, and to maximize all the rest. Knowing all the parameters for a given case it is then known in advance what the result would be, provided the effect is not seen.

To follow, we have to revisit the field of double beta decay and somewhat extend what we have said earlier. The double beta decay can obviously be of two types; the double neutron or the double proton decay (but, unlike single beta decay, not of both types at the same time). The double proton decay, depending on the Q-value, can be of three varieties: the double positron emission ($\beta^+\beta^+$), the positron emission - electron capture ($\beta^+EC$), and the double electron capture (EC,EC), each of the decays possibly having both the 2ν and 0ν component. The estimated (2ν+0ν)$\beta^+\beta^+$ decay half-lives are of the order of $10^{25}$ to $10^{30}$ years and it is not likely that these will be detected with our present-day sensitivities. Expected half-lives for the 2ν$\beta^+$EC decay are generally more than four orders of magnitude lower than those for the 2ν$\beta^+\beta^+$ decay [44] and this narrowed the list to the isotopes from the region around the Z,N=28 closed shell, where majority of candidates for this kind of decay is situated. Somehow, this is also probably the optimum range of atomic numbers for this type of decay; neither too low, from the view of electron capture, nor too high, from the view of the positron decay. Additionally, systematic study of these decays might hopefully reveal possible influence of shell effects on the probability of these processes. The decays in question are:

$$^{50}_{24}Cr_{26} \to ^{50}_{22}Ti_{28}, \quad ^{54}_{26}Fe_{28} \to ^{54}_{24}Cr_{30}, \quad ^{58}_{28}Ni_{30} \to ^{58}_{26}Fe_{32}, \quad ^{64}_{30}Zn_{34} \to ^{64}_{28}Ni_{36} \qquad (15)$$

and it is seen that the first and last of them close the magic number 28, what may result in their enhancement as compared to the second and the third one, which destroy the same magic number configuration. General adopted strategy is in this respect perhaps best expressed in the review [61] and Ref. 54.

Each of the decays (15), however, has its peculiarities which we shall now briefly discuss. All the promising double beta cases, with no competing processes which would contribute to background, are the decays of otherwise stable even-even nuclei from the bottom of mass parabolas, with $0^+$ ground states in both the parent and daughter nuclei. The ground state of the intermediate odd-odd nucleus is then higher than the ground state of the parent nucleus and the nuclear matrix element of the double beta transition contains the sum over all the allowed virtual states of this intermediate nucleus. This is where the 2ν and the 0ν modes greatly differ. In the 2ν mode the isospin selection rule forbids Fermi transitions, for which ΔT=0, and the decay is realized by two consecutive GT transitions only, for which ΔT=0,1. The virtual intermediate states can thus be practically only the $1^+$ states. Higher spins contribute negligibly, for at low energies the four emitted fermions are practically only s-waves. In the 0ν mode the exchanged virtual neutrino is spatially limited to the volume of the nucleus and its momentum may be quite high (while that of real neutrinos is limited by the Q-value of the decay), so that contributions of virtual intermediate states other than $1^+$ now can not be neglected. Moreover, the Fermi transitions are here not forbidden. Nuclear matrix elements are thus in two cases not identical. The matrix element, which is in the case of the 0ν mode needed to extract the value of the effective neutrino mass, thus can not be simply taken over from the simpler 2ν case, and this is why the estimates of neutrino mass from half-life limits are strongly model dependent. This is then why it is difficult to predict the decay rates of the decays (15) more quantitatively and, in the end, why we do not discuss possible implications of these experimental results on neutrino mass.

Basic relevant data for the four decays (15) are schematically presented in Fig.7. On the basis of the heuristic qualitative arguments discussed above the decays may be expected to have the following properties:

1) $^{50}_{24}Cr_{26} \rightarrow ^{50}_{22}Ti_{28}$ :

Since to create a positron $2m_ec^2$ is needed, this Q-value allows for the $\beta^+$EC and EC,EC decay modes, while the intermediate spins allow for the 0ν mode only. Closing of the N=28 magic number configuration may enhance the transition. Small isotopic abundance however greatly reduces the sensitivity. If only positrons are detected ("weighing" method) this would determine the partial half-life for the $0\nu\beta^+$EC decay mode.

2) $^{54}_{26}Fe_{28} \rightarrow ^{54}_{24}Cr_{30}$ :

This is an interesting case. Small Q-value allows for the EC,EC decay mode only, while the intermediate spins forbid the 2ν mode. Since in this situation no real radiations are emitted in the decay the only possibility for the decay to proceed is via a virtual excited state of the daughter nucleus, with the emission of this excitation energy in the form of real electromagnetic radiation (or the internal conversion electron). At this energy, however, there is no excited state in the daughter nucleus and the radiation of this energy is otherwise non-existent. Now, if this virtual state is an $0^+$ state then the emitted radiation is E0 and the ensuing conversion electron is observable with negligible efficiency only. If it is $2^+$ the gamma-ray will then have an energy of some 670 keV (depending on the type of the electron captures in the parent nucleus) and will be nicely detectable. In this case we would thus have the $0^+ \rightarrow 2^+$ transition of the 0ν mode, which allows for an interesting interpretation. Namely, it has been ascertained that as long as the weak interaction is described by a gauge theory the neutrinoless decay always means that the neutrino is a massive Majorana particle, independently of the existence or non-existence of right-handed currents [16]. However, the detection of a $0^+ \rightarrow 2^+$ transition of the 0ν mode (e.g. to the first excited state of the daughter nucleus) would mean that there exists the right-handed admixture to the weak interaction. (Because of the conservation of angular momentum and parity the contribution from the mass mechanism vanishes for this transition, in a first approximation [27]). The detection of this gamma-ray would thus possibly indicate the existence of right-handed (here alien left-handed!) weak currents. Destroying the N=28 shell this particular decay may however be expected to be hindered, the small isotopic abundance further reducing the chances to be observed.

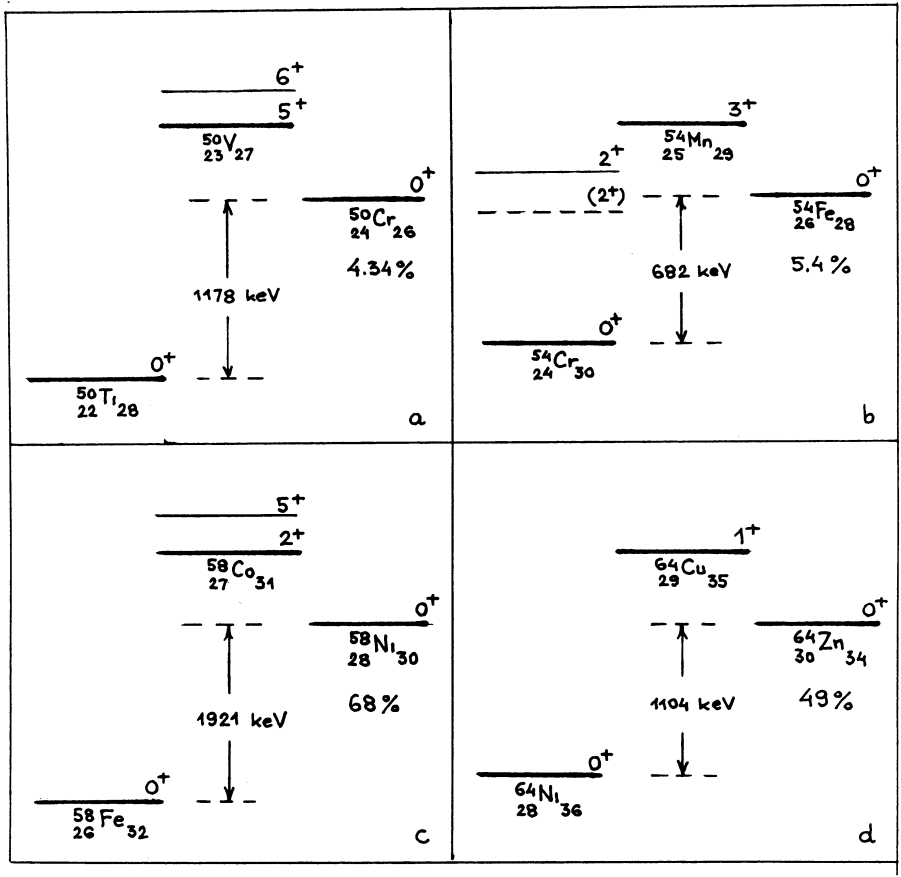

**Fig.7.** Relevant parts of the decay schemes of potential double proton decaying nuclei around Z,N=28.

3) $^{58}_{28}\text{Ni}_{30} \rightarrow ^{58}_{26}\text{Fe}_{32}$:

This decay is seen to be of the $0\nu(\beta^+\text{EC and EC,EC})$ type, of a rather high Q-value and the correspondingly favorable phase space factor. High isotopic abundance is perhaps compensated by the fact that the decay destroys the Z=28 shell. The decay has, however, been measured very well and it seemed unlikely that it would be possible to improve much upon the existing result.

4) $^{64}_{30}Zn_{34} \rightarrow ^{64}_{28}Ni_{36}$ :

This decay is perhaps the most promising one among those studied here. It is seen that it is of the $(2\nu+0\nu)(\beta^+EC$ and $EC,EC)$ type, with high isotopic abundance, and probably enhanced due to closing the Z=28 configuration. Detecting positrons would thus hopefully measure the $(2\nu+0\nu)(\beta^+EC)$ partial lifetime. This has decided this decay to be the first to be measured.

Finally, there is how these decays have been measured in Novi Sad.

The decays of $^{64}$Zn and $^{50}$Cr have been measured by practically the same technique. Experimental setup is schematically presented in Fig.8. Annihilation radiation is detected in coincidence by the 25% efficiency HpGe and the (3x3)" NaI(Tl) detectors. The two detectors are located in a central hole of the (9x9)" NaI(Tl) annulus, and the whole detector setup is situated in the 25 cm thick iron shield. Cylindrical sample of optimum thickness is sandwiched between the detectors. The fast-slow circuit minimizes the accidentals, and both the passive and active shielding is needed to minimize background induced by cosmic and environmental radiations. Iron shielding is for this purpose much better suited than the lead one, for the production of positrons by cosmic-ray muons depends on the square of the atomic number of the material divided by its mass number [29]. The 511 keV window is set on the stable Ge spectrum and the NaI spectrum is gated by the fast/slow signals. Coincidence efficiency is determined by calibrating with the $^{22}$Na source. Three different background measurements are performed: with Fe and Cu samples and with no sample at all. No statistically significant difference between these measurements was found. Average background intensity of the 511 keV line in the coincident spectrum, measured for 78 days, is 71(6)x10$^{-3}$ c/ks.

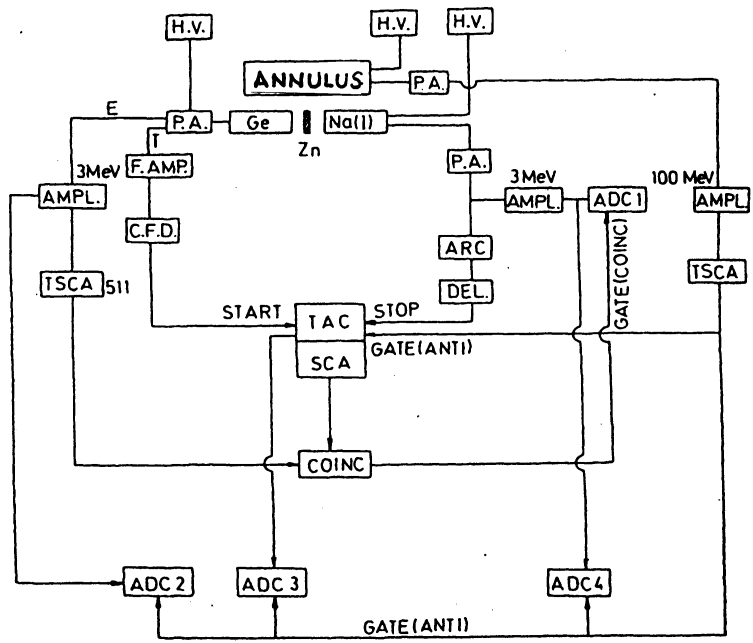

**Fig.8**. Experimental setup for the coincidence detection of positrons.

The $^{64}$Zn measurement lasted for 20 days and, at the 99.7% C.L. produced the non-zero result of 0.06(4) counts/ks over background (Fig.9) [63]. For the given experimental conditions this yields the half-life of:

$$T_{1/2} (2\nu+0\nu)(\beta^+EC) = 1.1(9) \times 10^{19} \text{ years}$$

at the 99.7% C.L. In spite of the favorable preliminary estimate this is a surprisingly low value. If it would be independently confirmed, it would be the first positive result for one $\beta^+$EC decay.

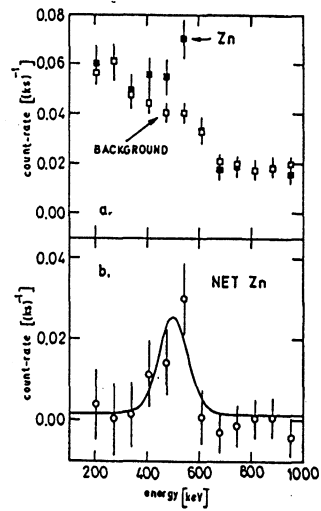

**Fig.9.** The intensity of the 511 keV line in the $(2\nu+0\nu)(\beta^+EC)$ decay of $^{64}$Zn.

The $^{50}$Cr measurement lasted for 30 days, during which, meeting the expectations, no significant difference between the sample and background counting rate was found. The resulting lower limit of the half-life at the 68% C.L. is then:

$$T_{1/2}(0\nu)(\beta^+EC) > 1.03 \times 10^{18} \text{ years}$$

this currently being the best estimate for this decay [56].

The measurement of the $^{54}$Fe decay is done in a completely different way. Advantage has been taken of the fact that all the routine measurements of background of the bare Ge detector (when used for direct gamma-ray spectroscopy of environmental samples) are performed within the 22 tones iron shielding, and that actually, among other things, represent the recordings of the radiations emitted by the iron itself. The total measurement time of 280 days could be accumulated in this way, the total amount of $^{54}$Fe in the shielding being 1.2 tons. However, taking the effects of self-absorption and detection efficiency into account, it turns out that only 32 grams of $^{54}$Fe has been looked upon with unit efficiency! In the sum spectrum no significant lines could be observed at the positions corresponding to KK, KL and LL electron captures, resulting in the lower limit on the half-life of:

$$T_{1/2}(0\nu)(KK) > 4.4 \times 10^{20} \text{ years}$$

at the 68% C.L. [64], [65]. Though some two orders of magnitude more stringent than the reported limits for the 0ν decay of $^{196}$Hg, this is presumably still far from reality. It is, however, not easy to substantially improve upon this result and it is bound to remain for some time the best experimental limit for this decay.

This is where we finish our review of the connections between Yugoslav and neutrino physics. The connections were few and the contributions modest, but Yugoslav physics may still be proud to have independently participated in this greatest of adventures. As it seems at the moment it is not likely that these connections will continue in the same spirit - home based research is loosing support and it is perhaps through international joint ventures that sometimes in the future will Yugoslav physics, this time probably with less initiative, again come into touch with the physics of the neutrino.

**Epilogue**

It remains for us to conclude this short review of the neutrino. Not much is left to be said. We can safely conclude that neutrino physics is alive and well. We have come a long way and now have a rather clear view of the road ahead - we are aware of what and how we are still to do. We have learned to keep an open mind when dealing with the neutrino and are ready for the surprises which we may encounter on that way.

Stimulating discussions with Prof. Djura Krmpotić during the laborious writing of this review, to whom the author owes much more than this, are greatly acknowledged. The thanks go to Prof. Istvan Bikit and Prof. Mića Pavićević for many years of keen friendship, devoted to the interests of the profession, without whom the second part of this review there would be not.

***

*Note added in proof (2002):* The appearance of two results stirred the neutrino community during the preparation of this manuscript for print.

Firstly, after 10 years of measurements, the Heidelberg-Moscow collaboration has announced (Mod.Phys.Lett. **A16**(2001)2409) the observation at the $3\sigma$ level of the neutrinoless double beta decay of Ge-76, with the half-life of 1.5e25 years, and the resulting Majorana neutrino mass of about 0.4 eV.

Secondly, in addition to the data from charged current neutrino interactions and neutrino elastic scattering within their heavy water detector, the SNO collaboration made use of the deuteron break-up events due to equal contributions of neutral current interactions of neutrinos of all flavors (the liberated neutrons being captured by the deuterons and the ensuing 6 MeV gamma-rays being detected by Cherenkov radiation). The rate of such events corroborate their earlier conclusions and suggest that the mu and tau neutrino flux is twice the electron neutrino flux, and that the total B-8 neutrino flux corresponds to that predicted by the SSM. Neutrino flavor oscillations thus appear to have been responsible all this time for the persistent solar neutrino problem (PRL nucl-ex/0204008, and www.sno.phy.queensu.ca). Moreover, they found the day-night asymmetry in neutrino interaction rates, suggesting matter enhanced contribution to neutrino oscillations (PRL nucl-ex/0204009).

We shall let the future to comment upon these results.